\documentclass[journal]{IEEEtran}
\IEEEoverridecommandlockouts
\usepackage{cite}
\usepackage{amsmath,amssymb,amsfonts}
\usepackage{algorithmic}
\usepackage[ruled,vlined]{algorithm2e}
\usepackage{graphicx}
\usepackage{subcaption}
\usepackage{textcomp}
\usepackage{cuted}   % in preamble
\usepackage{diagbox}
\usepackage{makecell}
\usepackage{threeparttable}
\usepackage{xcolor}
\usepackage{stfloats}
\allowdisplaybreaks
\usepackage[left=0.55in,right=0.55in,top=0.55in,bottom=0.55in]{geometry}
\usepackage[font=footnotesize,labelfont=footnotesize]{caption}
\def\BibTeX{{\rm B\kern-.05em{\sc i\kern-.025em b}\kern-.08em
    T\kern-.1667em\lower.7ex\hbox{E}\kern-.125emX}}

\begin{document}
\title{Joint Trajectory and Resource Optimization for Dual-aerial ARIS-assisted NOMA-TNT Networks}

\author{Vangara Saiprudhvi,~\IEEEmembership{Graduate Student Member,~IEEE},  Keshav Singh,~\IEEEmembership{Senior Member,~IEEE}, \\ Hariharan Subramaniyam,~\IEEEmembership{Senior Member,~IEEE}, and Chih-Peng Li,~\IEEEmembership{Fellow,~IEEE}
%\thanks{\hrule}

\thanks{V. Saiprudhvi is with the Institute of Communications Engineering, National Sun Yat-sen University, Kaohsiung 80424, Taiwan, and also with the Department of Communication Engineering, School of Electronics Engineering, Vellore Institute of Technology, Vellore, Tamilnadu, India (e-mail: saiprudhvi000@gmail.com).}

\thanks{Hariharan S. is with the Department of Communication Engineering, School of Electronics Engineering, Vellore Institute of Technology, Vellore, Tamilnadu, India (e-mail: shariharan@vit.ac.in).}

\thanks{K. Singh, and C.-P. Li are with the Institute of Communications Engineering, National Sun Yat-sen University, Kaohsiung 80424, Taiwan (e-mail: keshav.singh@mail.nsysu.edu.tw, cpli@mail.nsysu.edu.tw).}
\thanks{This work has been submitted to IEEE Transactions on Communications for possible publication.}

%\thanks{Yuanwei Liu is with the Department of Electrical and Electronic Engineering, The University of Hong Kong (HKU), Hong Kong (e-mail: yuanwei@hku.hk).}

\vspace{-2em}
}
        % <-this % stops a space
%\thanks{This paper was produced by the IEEE Publication Technology Group. They are in Piscataway, NJ.}% <-this % stops a space
%\thanks{Manuscript received April 19, 2021; revised August 16, 2021.}}



\maketitle

\begin{abstract}
Integrated terrestrial and non-terrestrial networks (ITNTNs) are envisioned as a key architectural paradigm for sixth-generation (6G) wireless systems, enabling seamless global connectivity. In this paper, we investigate a dual-aerial active reconfigurable intelligent surface (ARIS)-assisted non-orthogonal multiple access (NOMA)-based ITNTN, where a terrestrial base station (TBS) and a satellite base station (SAT) simultaneously serve terrestrial and satellite users with the assistance of an unmanned aerial vehicle (UAV)-mounted ARIS (UAV-ARIS) and a high-altitude platform (HAP)-mounted ARIS (HAP-ARIS). The users are multiplexed using power-domain NOMA with a predetermined successive interference cancellation (SIC) decoding order based on channel conditions.
We formulate an average sum-rate maximization problem by jointly optimizing the TBS and SAT transmit beamforming, the ARIS coefficients of the UAV-ARIS and HAP-ARIS, and the three-dimensional trajectories of the UAV and the HAP, subject to transmit power, unit-modulus, ARIS power budgets, and mobility constraints. The resulting optimization problem is highly non-convex due to the strong coupling among optimization variables, nonlinear SINR expressions, ARIS amplification factors, and trajectory-dependent channels.
To efficiently address this challenge, we propose a block coordinate descent (BCD)-based framework that decomposes the original problem into tractable subproblems. Specifically, transmit beamforming is optimized using the weighted minimum mean square error (WMMSE), ARIS phase-shifts are updated via a manifold-based Riemannian conjugate gradient (RCG) method, ARIS amplification factors are optimized using successive convex approximation (SCA), and first-order Taylor approximations for UAV and HAP trajectory optimization. The proposed algorithm is guaranteed to converge to a stationary point. 
The simulation results show that the proposed joint design achieves significant performance gains over the benchmark schemes. In particular, it achieves an average sum-rate improvement of approximately $8.44 \%$ compared to the passive RIS scheme under $\rho_{_U}^{max} = 2$ and $\rho_{_H}^{max} = 5$, highlighting the effectiveness of dual-aerial ARIS deployment and joint communication-mobility optimization in ITNTNs.
\end{abstract}
 
\section{Introduction}
Sixth-generation (6G) wireless networks are expected to significantly reshape future communication systems by enabling global, seamless, and reliable connectivity across heterogeneous environments \cite{11029408}. In this regard, integrated terrestrial and non-terrestrial networks (ITNTNs) have emerged as a key architectural paradigm, where terrestrial infrastructure is complemented by non-terrestrial platforms, such as satellites, high-altitude platforms (HAPs) and unmanned aerial vehicles (UAVs) \cite{10464935, 10082865,9992172}. By integrating these heterogeneous platforms, ITNTNs can provide extended coverage for remote and under-served areas, improve connectivity in dense urban environments, and improve network resilience during emergency and disaster scenarios \cite{10179219}. 
\par Despite these advantages, the heterogeneous architecture of ITNTNs introduces several technical challenges, including efficient resource allocation, coordinated spectrum sharing, interference mitigation, and cross-layer network management  \cite{9861699}. In addition, the long communication distance between aerial platforms and users often leads to severe propagation losses, which can significantly degrade the system performance. Addressing these issues is therefore essential to fully realize the potential benefits of ITNTN-enabled 6G networks. 

\subsection{Related Works}
\par Reconfigurable intelligent surfaces (RIS) have recently emerged as a promising technology for enhancing wireless communication performance by intelligently controlling the radio propagation environment \cite{9140329,9475160}. RIS consists of a large number of programmable reflecting elements that can adjust the phase shifts of incident signals to improve signal strength and coverage. Several studies have investigated RIS-assisted wireless networks to improve spectral efficiency, coverage, and energy efficiency in terrestrial networks \cite{8982186,8741198,9320618}. However, most of the existing work focuses on passive RIS, where the reflecting elements only adjust the phase of incident signals without amplification. As a result, passive RIS systems suffer from the \textit{double-fading effect}, which significantly limits their performance in long-distance communication scenarios. To overcome this limitation, active RIS (ARIS) has recently been proposed, where each reflecting element is equipped with active components capable of amplifying incident signals \cite{9377648,11456053,11390054}. Compared to passive RIS, ARIS can effectively compensate for severe path-loss and significantly improve communication performance, particularly in long-distance scenarios \cite{9998527}. Existing ARIS studies focus mainly on terrestrial networks, largely neglecting the advantages offered by aerial deployment and mobility-enabled channel optimization.
However, these works typically consider a single UAV-mounted passive RIS assisting terrestrial networks, and do not explore the potential advantages of multi-layer aerial architectures or ARIS technologies.
\par Recently, UAV-mounted RIS systems have attracted increasing attention due to their flexibility and adaptive deployment capabilities \cite{8959174,9599478}. By dynamically adjusting their positions, UAV-mounted RIS can improve channel conditions and enhance communication performance in complex environments \cite{9779521}. Several studies have investigated joint optimization of the UAV trajectory, RIS phase shifts, and transmit beamforming to improve network performance \cite{9684973,10319345}. However, these works typically consider a single UAV-mounted passive RIS assisting terrestrial networks, and do not explore the potential advantages of multi-layer aerial architectures or ARIS technologies.
\par In parallel, HAPs have been recognized as an effective platform for providing wide-area coverage due to their high altitude and large service footprint \cite{10186454}. HAP-assisted communication systems have been investigated as a promising solution to improve connectivity in remote and under-served regions within ITNTNs \cite{11121862}. However, the integration of RIS with HAP platforms has received limited attention, particularly in scenarios involving multi-layer aerial intelligent surfaces.
\par Furthermore, non-orthogonal multiple access (NOMA) has been widely studied as an efficient multi-user transmission technique capable of improving spectral efficiency by allowing multiple users to share the same time-frequency resources through power-domain multiplexing and successive interference cancellation (SIC) \cite{7676258,11408296}. The integration of NOMA with RIS-assisted networks has recently attracted significant interest due to its potential to further improve spectral efficiency and user connectivity \cite{9133094,9316920,11205889}. However, combining NOMA with multi-layer aerial RIS-assisted ITNTNs introduces additional design challenges due to the strong coupling among resource allocation, aerial mobility, and RIS configuration. 

\subsection{Motivation and Contributions}
\par Despite promising advances, several important challenges remain largely unexplored. First, most existing RIS-assisted communication systems consider only passive RIS or single aerial RIS deployment, without exploiting the potential benefits of multi-layer aerial RIS architectures. Second, the integration of ARIS with ITNTNs, which involve terrestrial and satellite networks, has received limited attention. Third, the joint optimization of transmit beamforming, ARIS coefficients, and aerial platform trajectories in NOMA-enabled ITNTNs results in a highly coupled and non-convex problem that remains largely unexplored in the literature.
\par Motivated by these observations, this paper proposes a dual-aerial ARIS-assisted NOMA-based ITNTN, where a UAV-ARIS and a HAP-ARIS simultaneously assist communications between a terrestrial base station (TBS), a satellite base station (SAT) and their associated users. By jointly optimizing transmit beamforming, ARIS coefficients (phase-shifts and amplification factors), and aerial trajectories, the proposed system aims to significantly improve the overall network spectral efficiency.\\
\begin{table*}[t]
\centering
\caption{Comparison with Existing RIS-Assisted Communication Works}
\label{tab:related_work_comparison}
\begin{tabular}{|c|c|c|c|c|c|}
\hline
\textbf{Reference} & \textbf{Network Type} & \textbf{RIS Type} & \textbf{Aerial Platform} & \textbf{NOMA} & \textbf{Trajectory Optimization} \\
\hline
\cite{8982186} & Terrestrial & Passive RIS & None & No & No \\
\hline
\cite{8741198} & Terrestrial & Passive RIS & None & No & No \\
\hline
\cite{9316920} & Terrestrial & Passive RIS & None & Yes & No \\
\hline
\cite{8959174} & Non-terrestrial & Passive RIS & UAV & No & Yes \\
\hline
\cite{9684973} & Non-terrestrial & Passive RIS & UAV & No & Yes \\
\hline
\cite{10538458} & Non-terrestrial & Active RIS & HAP & Yes & No\\ 
\hline
\textbf{This Work} & ITNTN & \textbf{Active RIS} & \textbf{UAV + HAP} & \textbf{Yes} & \textbf{Yes} \\
\hline
\end{tabular}
\end{table*}
The main contributions of this work are summarized as follows:
\begin{itemize}
    \item We propose a novel dual-aerial ARIS-assisted NOMA-based ITNTN architecture, in which a UAV-ARIS and a HAP-ARIS simultaneously assist communications between terrestrial and satellite networks that serve both terrestrial and satellite users. The proposed model captures heterogeneous propagation characteristics, three-dimensional (3D) mobility of aerial platforms, and multi-layer reflection effects enabled by dual RISs, thereby providing a general framework that includes single aerial RIS-assisted networks as special cases.

    \item We formulate an average sum-rate maximization problem by jointly optimizing the transmit beamforming vectors at the TBS and SAT, the ARIS coefficients (including phase-shifts and amplification factors) of both UAV-ARIS and HAP-ARIS, and the 3D trajectories of the UAV and HAP over multiple time slots.

    \item To address the resulting highly non-convex optimization problem, we develop an efficient block coordinate descent (BCD)-based algorithm that decomposes the original problem into multiple tractable subproblems. In each iteration, the transmit beamforming vectors are optimized using the weighted minimum mean square error (WMMSE), the RIS phase-shifts are updated using a manifold-based Riemannian conjugate gradient (RCG) method, the ARIS amplification factors are optimized via successive convex approximation (SCA), and the UAV/HAP trajectories are updated based on first-order Taylor approximations. 

    \item Extensive Monte Carlo simulations demonstrate that the proposed algorithm converges and achieves significant sum-rate improvements compared with benchmark schemes including no RIS and passive RIS. The results further validate the robustness of the proposed design in various system parameters, such as transmit power budget, RIS size, ARIS amplification factors, and flight duration.
\end{itemize}
\subsection{Organization and Notations}
\textit{Organization:} The remainder of this paper is organized as follows. Section \ref{ref:System Model} introduces the system model of the dual-aerial ARIS-assisted NOMA-based ITNTN. In section \ref{ref:Problem formulation}, we formulate the joint optimization problem. Then section \ref{ref:Proposed solution} presents the proposed BCD-based solution framework. The simulation results are provided in Section \ref{ref:Simulation results} to validate the effectiveness of the proposed design. Finally, Section \ref{ref:Conclusion} concludes the paper.
\par \textit{Notations:} Scalars, vectors, and matrices are denoted by $x$, $\mathbf{x}$, and $\mathbf{X}$, respectively. $\mathbb{C}^{M\times N}$ denotes the set of complex-valued matrices of size $M \!\times\! N$. $\mathbf{I}_N$ denotes the identity matrix $N\! \times\! N$. The operators $(\cdot)^T$ and $(\cdot)^H$ represent the transpose and the Hermitian transpose, respectively. The symbols $\circ$ and $\otimes$ denote the Hadamard product and the Kronecker product, respectively. The complex circularly symmetric Gaussian distribution with mean $\mu$ and variance $\sigma^2$ is denoted by $\mathcal{CN}(\mu,\sigma^2)$. The absolute value of a scalar and the Euclidean norm of a vector are denoted by $|\cdot|$ and $\|\cdot\|$, respectively. $\mathrm{diag}(\cdot)$ denotes a diagonal matrix. $\Re\{\cdot\}$ denotes the real part of a complex number, $\iota \triangleq \sqrt{-1}$ denotes the imaginary unit, and $\mathbb{E}[\cdot]$ denotes the expectation operator.
\section{System Model}\label{ref:System Model}
\begin{figure}[t]
    \centering
    \includegraphics[scale=0.1]{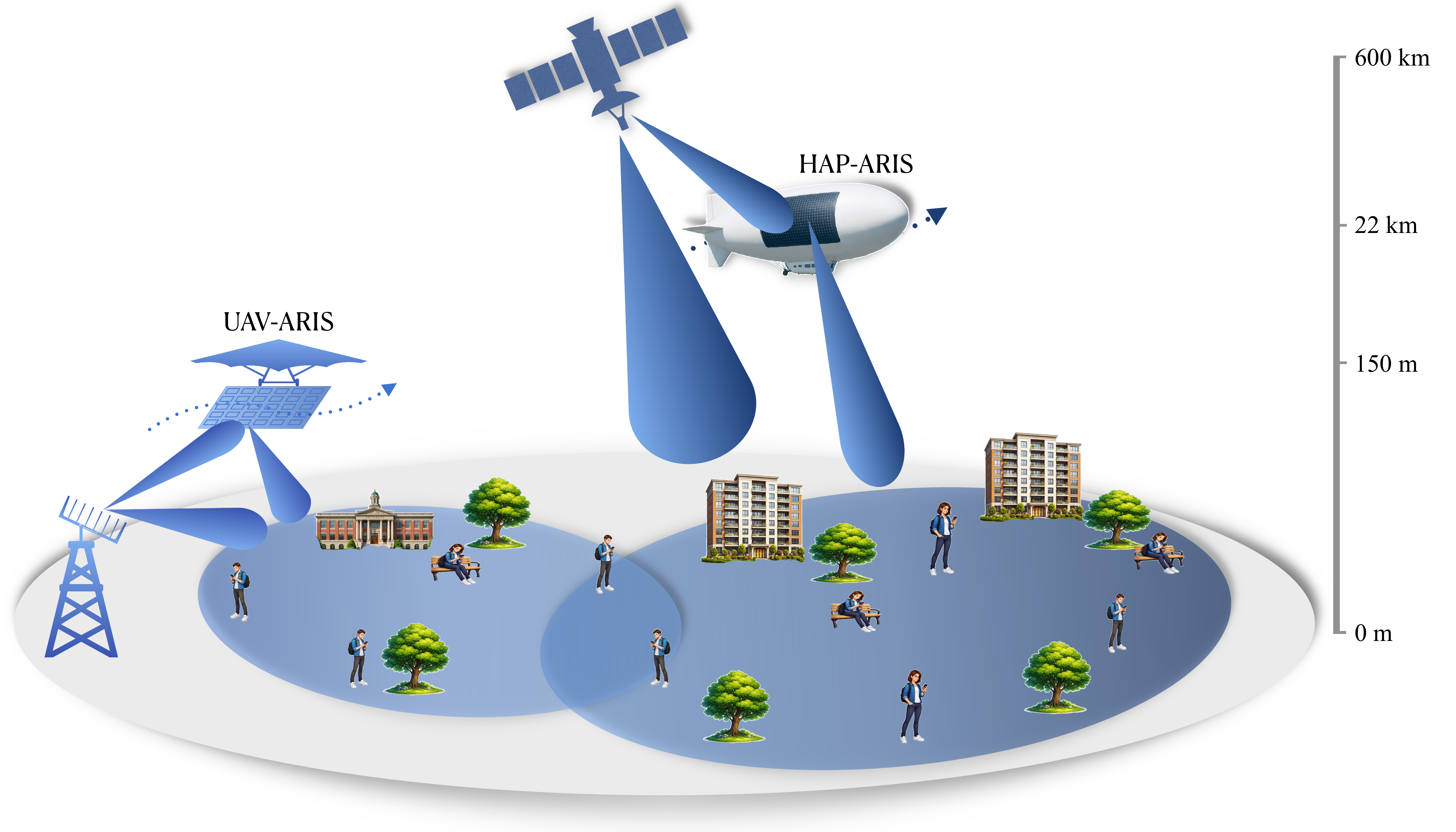}
    \caption{System model of dual aerial ARIS-assisted NOMA-based ITNTN.}
    \label{SystemModel}
\end{figure}
We consider an ITNTN in which the TBS and SAT simultaneously serve terrestrial and satellite users, respectively, with the assistance of a UAV-ARIS and HAP-ARIS.
The TBS is equipped with a uniform linear array (ULA) consisting of $M_b$ antennas with an inter-element spacing of $d = \lambda/2$, where $\lambda=c/f_c$, $c$ denotes the speed of light and $f_c$ is the carrier frequency of terrestrial link. The TBS serves $K$ terrestrial users indexed by $k \in \{1,\dots,K\}$. 
The SAT employs a uniform planar array (UPA) with $M_s$ antennas and an inter-element spacing of $\check{d} = \check{\lambda}/2$, where $\check{\lambda} = c/\check{f}_c$ and $\check{f}_c$ denote the carrier frequency of the satellite link. It serves $L$ satellite users indexed by $l \in \{1,\dots,L\}$. 
Both UAV-ARIS and HAP-ARIS are equipped with UPAs comprising $N_U = N_x \times N_y$ and $N_H = N_{h,x} \times N_{h,y}$ reflecting elements, respectively. These platforms are mobile and their 3D trajectories are jointly optimized over $N$ equal time slots, each of duration $\delta = T/N$. 
The system architecture is illustrated in Fig.~\ref{SystemModel}.
\subsection{Channel Model}
In this subsection, we present the channel models for all communication links in the considered ITNTN. All channels are assumed quasi-static within each time slot and vary independently across different slots. 
\subsubsection{TBS-to-terrestrial user $k$ links}
The positions of the TBS and the terrestrial user $k$ are given by $\mathbf{q}_{_{TBS}} \!=\! (x_{_{TBS}}, y_{_{TBS}}, 0)$ and $\mathbf{q}_{_k} \!=\! (x_{_k}, y_{_k}, 0)$, respectively. The 3D position of UAV-ARIS at the time slot $n$ is denoted by $\mathbf{q}_{_U}[n]\!=\!\left(x_{_U}[n],y_{_U}[n],z_{_U}[n]\right)$. 
\par The direct TBS-to-user $k$ channel $\mathbf{h}_{b,k} \in \mathbb{C}^{1\times M_b}$ follows a Rician fading model and is given by
\begin{align}\label{TBS-to-k}
    \mathbf{h}_{b,k} \!=\! \sqrt{\beta_0d_{b,k}^{-\eta_{_{b,k}}}}\!\left(\!\sqrt{\tfrac{\kappa_{_{b,k}}}{1+\kappa_{_{b,k}}}}\mathbf{\check{h}}_{b,k}\!+\!\sqrt{\tfrac{1}{1+\kappa_{_{b,k}}}}\mathbf{\breve{h}}_{b,k}\!\right),
\end{align}
where $\kappa_{_{b,k}}$ denotes the Rician factor, $\beta_0$ is the reference path-loss at a distance of $1$ m, and $\eta_{_{b,k}}$ is the path-loss exponent. The distance between the TBS and the terrestrial user $k$ is $d_{b,k} \!=\! \|\mathbf{q}_{_{TBS}}\!-\!\mathbf{q}_{_k}\|$. The non-line-of-sight (NLoS) component $\mathbf{\breve{h}}_{b,k}$ has i.i.d entries following $\mathcal{CN}(0,1)$ \cite{11155890}. The line-of-sight (LoS) component is expressed as $\mathbf{\check{h}}_{b,k} = \mathbf{a}^H_{_{TBS}}(\psi_{b}^k,\phi_{b}^k)$, where $\psi_b^k$ and $\phi_{b}^k$ denote the azimuth and elevation angles of user $k$, respectively.
The ULA steering vector at the TBS $a_{_{TBS}}(\psi,\phi)$ is given by
\begin{align}
    \mathbf{a}_{_{TBS}}(\psi,\phi) \!=\! \tfrac{1}{\sqrt{M_b}}\Big[ 1, &e^{-\iota2\pi\frac{d}{\lambda}\cos\phi\cos\psi},
    \dots,\nonumber\\&\hspace{2em} e^{-\iota2\pi\frac{(M_b-1)d}{\lambda}\cos\phi\cos\psi}
    \Big]^T,
\end{align}
where $\psi$ and $\phi$ denote azimuth and elevation angles, respectively.
\par The TBS-to-UAV-ARIS channel $\mathbf{H}_{b,U}\in \mathbb{C}^{N_U\times M_b}$ and the UAV-ARIS to-user $k$ channel $\mathbf{h}_{U,k}\in \mathbb{C}^{N_U\times 1}$ follow Rician fading models and are expressed as
\begin{align}
    \mathbf{H}_{b,U} \!&=\!\! \sqrt{\beta_0d_{b,U}^{-\eta_{_{b,U}}}}\!\!\left(\!\!\sqrt{\tfrac{\kappa_{_{b,U}}}{1+\kappa_{_{b,U}}}}\mathbf{\check{H}}_{b,U}\!+\!\sqrt{\tfrac{1}{1+\kappa_{_{b,U}}}}\mathbf{\breve{H}}_{b,U}\!\!\right)\!,\\
    \mathbf{h}_{U,k} \!&=\!\! \sqrt{\beta_0d_{U,k}^{-\eta_{_{U,k}}}}\!\!\left(\!\!\sqrt{\tfrac{\kappa_{_{U,k}}}{1+\kappa_{_{U,k}}}}\mathbf{\check{h}}_{U,k}\!+\!\sqrt{\tfrac{1}{1+\kappa_{_{U,k}}}}\mathbf{\breve{h}}_{U,k}\!\!\right)\!,
\end{align}
where $d_{b,U}=\|\mathbf{q}_{_{TBS}}-\mathbf{q}_{_U}[n]\|$ and $d_{U,k}=\|\mathbf{q}_{_{U}}[n]-\mathbf{q}_{_k}\|$ denote the distance from the TBS to the UAV-ARIS and from the UAV-ARIS to the terrestrial user $k$, respectively. The parameters $\eta_{_{b,U}}$ and $\eta_{_{U,k}}$ are the path loss exponents, while $\kappa_{_{b,U}}$ and $\kappa_{_{U,k}}$ denote the Rician factors. The NLoS components $\mathbf{\breve{H}}_{b,U}$ and $\mathbf{\breve{h}}_{U,k}$ consist of i.i.d. entries following $\mathcal{CN}(0,1)$ \cite{11155890}. The LoS components are given by $\mathbf{\check{H}}_{b,U} = \mathbf{a}_{_U}(\psi_{_U},\phi_{_U})\mathbf{a}_{_{TBS}}^H(\psi_{b}^U,\phi_{b}^U)$ and $\mathbf{\check{h}}_{U,k} = \mathbf{a}_{_U}(\psi_{_U}^k,\phi_{_U}^k)$, where $(\psi_{_U},\phi_{_U})$ denote the azimuth and elevation angles of arrival (AoA) at the UAV-ARIS from the TBS, $(\psi_{b}^U,\phi_{b}^U)$ denote the angle of departure (AoD) at the TBS and $(\psi_{_U}^k,\phi_{_U}^k)$ denote the AoD from the UAV-ARIS to the terrestrial user $k$.
The UPA steering vector at the UAV-ARIS is given by
\begin{align}
    \mathbf{a}_{_U}(\psi,\phi) = \tfrac{1}{\sqrt{N_U}}\big(\mathbf{a}_x(\psi,\phi)\otimes\mathbf{a}_y(\psi,\phi)\big),
\end{align}
where
\begin{align}
\mathbf{a}_{x}(\psi,\phi)\! &=\! \tfrac{1}{\sqrt{N_x}}\Big[\! 1, e^{-\iota2\pi\frac{d}{\lambda}\cos\phi\cos \psi},\nonumber\\& \hspace{5em}
\dots, e^{-\iota2\pi\frac{(N_x-1)d}{\lambda}\cos\phi\cos \psi}
\!\Big]^T\!, \\
\mathbf{a}_{y}(\psi,\phi)\! &=\! \tfrac{1}{\sqrt{N_y}}\Big[\! 1, e^{-\iota2\pi\frac{d}{\lambda}\cos\phi\sin \psi},\nonumber\\& \hspace{5em} 
\dots, e^{-\iota2\pi\frac{(N_y-1)d}{\lambda}\cos\phi\sin \psi}
\!\Big]^T.
\end{align}
\par Finally, the effective channel between TBS and the terrestrial user $k$ at the time slot $n$ is expressed as
\begin{align}
    \mathbf{h}_{b,k}^{\mathrm{eff}}[n] = \mathbf{h}_{b,k}+\mathbf{h}_{U,k}^H[n]\boldsymbol{\Theta}_U[n]\mathbf{H}_{b,U}[n],
\end{align} 
where $\boldsymbol{\Theta}_U[n]\!\triangleq\! \mathrm{diag}\big(\varphi_{_{U,1}}[n],\dots,\varphi_{_{U,N_U}}[n]\big) \!\in\! \mathbb{C}^{N_U\times N_U}$ is the UAV-ARIS coefficient matrix. The reflection coefficient of the $r$-th ARIS element is given by \cite{9998527}
\begin{align}
    \varphi_{_{U,r}}[n]=\sqrt{\rho_{_{U,r}}[n]}e^{\iota\theta_{{U,r}}[n]},
\end{align}
where $\theta_{{U,r}}[n]\in [0,2\pi)$ and $\rho_{_{U,r}}$ denote the phase shift and amplification factor, respectively, with $r\in\{1,\dots,N_U\}$.
\subsubsection{SAT-to-satellite user $l$ links} The positions of the SAT and satellite user $l$ are given by $\mathbf{q}_{_{SAT}} = (x_{_{SAT}}, y_{_{SAT}}, H_{{SAT}})$ and $\mathbf{q}_{_l} = (x_{_l}, y_{_l}, 0)$, where $H_{SAT}$ is the altitude of the SAT. The 3D position of HAP-ARIS at the time slot $n$ is denoted by $\mathbf{q}_{_H}=\left(x_{_H}[n],y_{_H}[n],z_{_H}[n]\right)$. 
The direct SAT-to-user $l$ channel $\mathbf{h}_{s,l} \in \mathbb{C}^{1 \times M_s}$ follows a Rician fading model and is given by
\begin{align}\label{SAT-to-l}
    \mathbf{h}_{s,l} \!=\! \sqrt{\mathcal{L}_{s,l}}\!\left(\!\sqrt{\tfrac{\kappa_{_{s,l}}}{1+\kappa_{_{s,l}}}}\mathbf{\check{h}}_{s,l}\!+\!\sqrt{\tfrac{1}{1+\kappa_{_{s,l}}}}\mathbf{\breve{h}}_{s,l}\!\right),
\end{align}
where $\kappa_{_{s,l}}$ denotes the Rician factor and $\mathcal{L}_{s,l}$ represents the large-scale fading component, which is given by
\begin{align}
    \mathcal{L}_{s,l} \!=\! \left(\!\!\tfrac{\check{\lambda}}{4\pi d_{s,l}}\!\!\right)^2G_{s}G_{_l}r_s,
\end{align}
where $d_{s,l}\!=\!\|\mathbf{q}_{_{SAT}}\!-\!\mathbf{q}_{_l}\|$ denotes the distance between the SAT and the satellite user $l$, $G_s$ is the satellite antenna gain and $G_{_l}$ denotes the user's receive antenna gain. The rain attenuation coefficient $r_s$ follows a log-normal distribution, i.e., $\ln(r_s^{dB})\!\sim\!\mathcal{N}(\mu_r,\sigma_r^2)$ \cite{3GPP_TR_38_901}.
The NLoS component $\mathbf{\breve{h}}_{s,l}$ has i.i.d. entries following $\mathcal{CN}(0,1)$ \cite{11155890}.
The SAT UPA responses are given by 
\begin{align}
    \mathbf{a}_{_{SAT}}(\psi,\phi) = \tfrac{1}{\sqrt{M_s}}\big(\mathbf{a}_{_{SAT}}^x(\psi,\phi)\otimes\mathbf{a}_{_{SAT}}^y(\psi,\phi)\big),
\end{align}
where
\begin{align}
&\mathbf{a}_{_{SAT}}^{x}(\psi,\phi)
= \tfrac{1}{\sqrt{M_{s,x}}}
\Big[
 1,
e^{-\iota2\pi\frac{\check{d}_x}{\check{\lambda}}\cos\phi\cos\psi},  \nonumber\\
& \hspace{9em}\dots,
e^{-\iota2\pi\frac{(M_{s,x}-1)\check{d}_x}{\check{\lambda}}
     \cos\phi\cos\psi}
\Big]^{T},\\
&\mathbf{a}_{_{SAT}}^{y}(\psi,\phi)
= \tfrac{1}{\sqrt{M_{s,y}}}
\Big[
 1,
e^{-\iota2\pi\frac{\check{d}_y}{\check{\lambda}}\cos\phi\sin\psi},\nonumber\\ &\hspace{9em} \dots,e^{-\iota2\pi\frac{(M_{s,y}-1)\check{d}_y}{\check{\lambda}}\cos\phi\sin\psi}
\Big]^{T},
\end{align}
Here, $M_{s,x}$ and $M_{s,y}$ are the numbers of SAT UPA elements along the $x$ and $y$-axes, respectively, and $\check{d}_x$ and $\check{d}_y$ are the corresponding antenna spacings. Let $\psi_{_{s,l}}$ and $\phi_{_{s,l}}$ denote the azimuth and elevation angles of user $l$, respectively. The LoS component corresponds to the SAT UPA steering vector, given by 
$\mathbf{\check{h}}_{s,l} = \mathbf{a}_{_{SAT}}^H(\psi_{_{s,l}},\phi_{_{s,l}})$.
\par The channel between SAT-to-HAP-ARIS $\mathbf{H}_{s,H}\in \mathbb{C}^{N_H\times M_s}$ and HAP-ARIS-to-user $l$ $\mathbf{h}_{H,l}\in\mathbb{C}^{N_H\times1}$ also follows Rician fading models and are given by
\begin{align}
    \mathbf{H}_{s,H} \!&=\! \sqrt{\mathcal{L}_{s,H}}\!\left(\!\!\sqrt{\tfrac{\kappa_{_{s,H}}}{1\!+\!\kappa_{_{s,H}}}}\mathbf{\check{H}}_{s,H}\!+\!\sqrt{\tfrac{1}{1\!+\!\kappa_{_{s,H}}}}\mathbf{\breve{H}}_{s,H}\!\!\right),\\
    \mathbf{h}_{H,l} \!&=\! \sqrt{\mathcal{L}_{H,l}}\!\left(\!\sqrt{\tfrac{\kappa_{_{H,l}}}{1+\kappa_{_{H,l}}}}\mathbf{\check{h}}_{H,l}\!+\!\sqrt{\tfrac{1}{1+\kappa_{_{H,l}}}}\mathbf{\breve{h}}_{H,l}\!\right),
\end{align}
where $\mathcal{L}_{s,H}$ and $\mathcal{L}_{H,l}$ denotes the large-scale fading components, which are given by
\begin{align}
    \mathcal{L}_{s,H} \!=\! \left(\!\!\tfrac{\check{\lambda}}{4\pi d_{s,H}}\!\!\right)^2\!\!G_{s}G_{H},\;\;
    \mathcal{L}_{H,l} \!=\! \left(\!\!\tfrac{\check{\lambda}}{4\pi d_{H,l}}\!\!\right)^2\!\!G_{H}G_{_l}r_s,
\end{align}
with $d_{s,H}=\|\mathbf{q}_{_{SAT}}-\mathbf{q}_{_H}[n]\|$ and $d_{H,l}=\|\mathbf{q}_{_{H}}[n]-\mathbf{q}_{_l}\|$ denote the distance from SAT to HAP-ARIS and HAP-ARIS to the satellite user $l$, respectively.
The parameters $\kappa_{_{s,H}}$ and $\kappa_{_{H,l}}$ are Rician factors. The NLoS component $\mathbf{\breve{H}}_{s,H}$ and $\mathbf{\breve{h}}_{H,l}$ consists of i.i.d. entries following $\mathcal{CN}(0,1)$ \cite{11155890}.
The HAP-ARIS UPA steering vector is given by 
\begin{align}
    \mathbf{a}_{_H}(\psi,\phi) = \tfrac{1}{\sqrt{N_H}}\big(\mathbf{a}_{_H}^x(\psi,\phi)\otimes\mathbf{a}_{_H}^y(\psi,\phi)\big),
\end{align}
where
\begin{align}
&\mathbf{a}_{_H}^{x}(\psi,\phi)
= \tfrac{1}{\sqrt{N_{h,x}}}
\Big[
 1,
e^{-\iota2\pi\frac{\check{d}_{h,x}}{\check{\lambda}}\cos\phi\cos\psi},  \nonumber\\
&\hspace{5em} \dots,\;
e^{-\iota2\pi\frac{(N_{h,x}-1)\check{d}_{h,x}}{\check{\lambda}}
     \cos\phi\cos\psi}
\Big]^{T},\\
&\mathbf{a}_{_H}^{y}(\psi,\phi)
= \tfrac{1}{\sqrt{N_{h,y}}}
\Big[
 1,
e^{-\iota2\pi\frac{\check{d}_{h,y}}{\check{\lambda}}\cos\phi\sin\psi},\nonumber\\ &\hspace{5em} \dots,e^{-\iota2\pi\frac{(N_{h,y}-1)\check{d}_{h,y}}{\check{\lambda}}\cos\phi\sin\psi}
\Big]^{T},
\end{align}
Here, \!$N_{h,x}$ \!and \!$N_{h,y}$\! are the numbers of HAP-ARIS UPA elements along the $x$ and $y$-axes, respectively, and $\check{d}_{h,x}$ and $\check{d}_{h,y}$ are the corresponding antenna spacings. Let $\psi_{_H}$ and $\phi_{_H}$ denote the azimuth and elevation AoA at the HAP-ARIS from the SAT, and $\psi_{s,h}$ and $\phi_{s,h}$ denote the AoD at the SAT.
The LoS components of SAT-to-HAP-ARIS and HAP-ARIS-to-user $l$ are given by 
$\check{\mathbf{H}}_{s,H} = \mathbf{a}_{_H}(\psi_{_H},\phi_{_H})\mathbf{a}_{_{SAT}}^H(\psi_{s,h},\phi_{s,h})$
and
$\check{\mathbf{h}}_{H,l}
= \mathbf{a}_{_H}(\psi_{_H}^l,\phi_{_H}^l)$.
\par Finally, the effective channel between the SAT and the satellite user $l$ at the time slot $n$ is expressed as
\begin{align}
    \mathbf{h}_{s,l}^{\mathrm{eff}}[n] = \mathbf{h}_{s,l}+\mathbf{h}^H_{H,l}\boldsymbol{\Theta}_H[n]\mathbf{H}_{s,H}[n],
\end{align}
where $\boldsymbol{\Theta}_H[n]\!\triangleq\! \mathrm{diag}\big(\varphi_{_{H,1}}[n],\dots,\varphi_{_{H,N_H}}[n]\big) \!\in\! \mathbb{C}^{N_H\times N_H}$ is the HAP-ARIS coefficient matrix. The reflection coefficient of the $\acute{r}$-th ARIS element is given by \cite{9998527}
\begin{align}
    \varphi_{_{H,\acute{r}}}[n]=\sqrt{\rho_{_{H,\acute{r}}}[n]}e^{\iota\theta_{{H,\acute{r}}}[n]},
\end{align}
where $\theta_{{H,\acute{r}}}[n]\in [0,2\pi)$ and $\rho_{_{H,\acute{r}}}$ denote the phase shift and amplification factor, respectively, with $\acute{r}\in\{1,\dots,N_H\}$.
\subsection{Input-output Relations}
The TBS adopts power-domain NOMA to simultaneously serve $K$ terrestrial users over the same time-frequency resource block. The transmitted signal from the TBS is given by
\begin{align}
    \mathbf{x}_b[n] = \sum\nolimits_{k=1}^K\mathbf{v}_{b,k}[n]s_{b,k}[n],
\end{align}
where $\mathbf{v}_{b,k}[n]=\mathbf{w}_{b,k}[n]\sqrt{p_{b,k}[n]}\!\in\!\mathbb{C}^{M_b\times1}$ denotes the effective beamforming vector for the terrestrial user $k$, $p_{b,k}[n]$ is the power allocation coefficient, and $s_{b,k}[n]$ is the information symbol satisfying $\mathbb{E}[|s_{b,k}[n]|^2]\!=\!1$. 
Similarly, the SAT employs power-domain NOMA to serve \!$L$\! satellite users, and its transmitted signal is expressed as
\begin{align}
    \mathbf{x}_s[n] = \sum\nolimits_{l=1}^L\mathbf{v}_{s,l}[n]s_{s,l}[n],
\end{align}
where $\mathbf{v}_{s,l}[n]=\mathbf{w}_{s,l}[n]\sqrt{p_{s,l}[n]}\in\mathbb{C}^{M_s\times1}$ denotes the effective beamforming vector for the satellite user $l$, $p_{s,l}[n]$ is the power allocation coefficient, and $s_{s,l}[n]$ satisfies $\mathbb{E}[|s_{s,l}[n]|^2]\!=\!1$. 
\\
At time slot $\!n$, the received signal at terrestrial user \!$k$\! is given by
\begin{align}
y_{_k}[n] &= \underbrace{\mathbf{h}_{b,k}^{\mathrm{eff}}[n]\mathbf{v}_{b,k}[n]s_{b,k}[n]}_{\text{desired signal}} \nonumber\\&\hspace{1.5em} +\!\underbrace{\sum\nolimits_{i=1,i\neq k}^K\mathbf{h}_{b,k}^{\mathrm{eff}}[n]\mathbf{v}_{b,i}[n]s_{b,i}[n]}_{\text{intra-TBS interference }} \nonumber\\
&\hspace{2.5em}+\underbrace{\sum\nolimits_{l=1}^{L}\mathbf{h}_{s,k}^{\mathrm{eff}}[n]\mathbf{v}_{s,l}[n]s_{s,l}[n]}_{\text{SAT-to-user $k$ interference}\;}\nonumber\\
&\hspace{3.5em}+\mathbf{h}_{U,k}^H[n]\boldsymbol{\Theta}_U[n]\mathbf{n}_U  +n_k[n],
\end{align}
where $n_k[n]\!\sim\!\mathcal{CN}(0,\sigma_k^2)$ denotes additive white Gaussian noise (AWGN) at terrestrial user $k$ and $\mathbf{n}_U\!\sim\! \mathcal{CN}(\mathbf{0},\sigma^2_U\mathbf{I})$ represents the thermal noise introduced by the active elements of UAV-ARIS, where $\sigma^2_U$ is the noise variance of each  UAV-ARIS element. For terrestrial user $k$, the interference from the SAT consists of a direct SAT-to-user $k$ link and a cascaded SAT-to-HAP-ARIS-to-user $k$ link. The corresponding effective cross-tier channel is defined as
\begin{align}
\mathbf{h}_{s,k}^{\mathrm{eff}}[n] = \mathbf{h}_{s,k} + \mathbf{h}_{H,k}^{H}[n]
\boldsymbol{\Theta}_{H}[n]\mathbf{H}_{s,H}[n],
\end{align}
where $\mathbf{h}_{s,k}$ denotes the direct SAT-to-user $k$ channel, and $\mathbf{h}_{H,k}$ is the HAP-ARIS-to-user $k$ channel. Both channels are modeled using the same Rician fading model as in \eqref{SAT-to-l}, with appropriate large-scale fading components.
%%%%%%%%%%%% Bottom %%%%%%%%%%%%%%%%%%
\begin{figure*}[!b]
\hrulefill
% \noindent\textit{The downlink SINR expressions for the terrestrial and satellite NOMA users are given as}
\begin{align}
    &\gamma_{k\to i}[n] \!=\! \frac{\big|\mathbf{h}_{b,k}^{\mathrm{eff}}[n]\mathbf{v}_{b,i}[n]\big|^2 }{\sum\nolimits_{m = i+1}^K\big|\mathbf{h}_{b,k}^{\mathrm{eff}}[n]\mathbf{v}_{b,m}[n]\big|^2+\sum\nolimits_{l=1}^L\big|\mathbf{h}_{s,k}^{\mathrm{eff}}[n]\mathbf{v}_{s,l}[n]\big|^2+\|\mathbf{h}^H_{U,k}[n]\boldsymbol{\Theta}_U[n]\|^2\sigma^2_U+\sigma_k^2}, \;\; \forall \;1\!\leq\! i\!<\!k\!\leq \!K, \label{TBS_SINR_i}\\
    &\gamma_{k}[n] = \frac{\big|\mathbf{h}_{b,k}^{\mathrm{eff}}[n]\mathbf{v}_{b,k}[n]\big|^2}{\sum\nolimits_{m = k+1}^K\big|\mathbf{h}_{b,k}^{\mathrm{eff}}[n]\mathbf{v}_{b,m}[n]\big|^2 +\sum\nolimits_{l=1}^L\big|\mathbf{h}_{s,k}^{\mathrm{eff}}[n]\mathbf{v}_{s,l}[n]\big|^2+\|\mathbf{h}^H_{U,k}[n]\boldsymbol{\Theta}_U[n]\|^2\sigma^2_U+\sigma_k^2}, \quad \forall \;1\leq k\leq K\label{TBS_SINR_k},\\
    &\gamma_{l\to j}[n] \!=\! \frac{\big|\mathbf{h}_{s,l}^{\mathrm{eff}}[n]\mathbf{v}_{s,j}[n]\big|^2}{\sum\nolimits_{m = j+1}^L\big|\mathbf{h}_{s,l}^{\mathrm{eff}}[n]\mathbf{v}_{s,m}[n]\big|^2+\sum\nolimits_{k=1}^K\big|\mathbf{h}_{b,l}^{\mathrm{eff}}[n]\mathbf{v}_{b,k}[n]\big|^2+\|\mathbf{h}^H_{H,l}[n]\boldsymbol{\Theta}_H[n]\|^2\sigma^2_H+\sigma_l^2},\;\;\forall \;1\!\leq\! j\!<\!l\!\leq \!L, \label{SAT_SINR_j}\\
    &\gamma_{l}[n] = \frac{\big|\mathbf{h}_{s,l}^{\mathrm{eff}}[n]\mathbf{v}_{s,l}[n]\big|^2}{\sum\nolimits_{m = l+1}^L\big|\mathbf{h}_{s,l}^{\mathrm{eff}}[n]\mathbf{v}_{s,m}[n]\big|^2+\sum\nolimits_{k=1}^K\big|\mathbf{h}_{b,l}^{\mathrm{eff}}[n]\mathbf{v}_{b,k}[n]\big|^2+\|\mathbf{h}^H_{H,l}[n]\boldsymbol{\Theta}_H[n]\|^2\sigma^2_H+\sigma_l^2}, \quad \forall \;1\leq l\leq L\label{SAT_SINR_l}.
\end{align}
\end{figure*}
%%%%%%%%%%%%%%%%%%% bottom %%%%%%%%%%%%
Similarly, the received signal at satellite user $l$ during time slot $n$ is given by
\begin{align}
y_{_l}[n] &= \underbrace{\mathbf{h}_{s,l}^{\mathrm{eff}}[n]\mathbf{v}_{s,l}[n]s_{s,l}[n]}_{\text{desired signal}} \nonumber\\
&\hspace{1.5em}+\!\underbrace{\sum\nolimits_{j=1,j\neq l}^L\mathbf{h}_{s,l}^{\mathrm{eff}}[n]\mathbf{v}_{s,j}[n]s_{s,j}[n]}_{\text{intra-SAT interference }} \nonumber\\
&\hspace{2.5em}+\underbrace{\sum\nolimits_{k=1}^{K}\mathbf{h}_{b,l}^{\mathrm{eff}}[n]\mathbf{v}_{b,k}[n]s_{b,k}[n]}_{\text{TBS-to-user $l$ interference}\;} \nonumber\\
&\hspace{3.5em}+\mathbf{h}_{H,l}^H[n]\boldsymbol{\Theta}_H[n]\mathbf{n}_H  +n_l[n],
\end{align}
where $n_{l}[n]\!\sim\!\mathcal{CN}(0,\sigma_l^2)$ denotes the AWGN at satellite user $l$ and $\mathbf{n}_H\!\sim\! \mathcal{CN}(\mathbf{0},\sigma^2_H\mathbf{I})$ represents the thermal noise introduced by the active elements of HAP-ARIS, where $\sigma^2_H$ is the noise variance of each HAP-ARIS element. For satellite user $l$, the interference from the TBS consists of a direct TBS-to-user $l$ link and a cascaded TBS-to-UAV-ARIS-to-user $l$ link. The effective cross-tier channel is defined as
\begin{align}
\mathbf{h}_{b,l}^{\mathrm{eff}}[n] = \mathbf{h}_{b,l} + \mathbf{h}_{U,l}^{H}[n]
\boldsymbol{\Theta}_{U}[n]\mathbf{H}_{b,U}[n],
\end{align}
where $\mathbf{h}_{b,l}$ denotes the direct TBS-to-user $l$ channel, and $\mathbf{h}_{U,l}$ is the UAV-ARIS-to-user $l$ channel. Both channels are modeled using the same Rician fading model as in \eqref{TBS-to-k}, with appropriate large-scale fading components.\\
Without loss of generality, terrestrial users are indexed in ascending order of their effective channel gains as \cite{9316920}
\begin{align}
    \|\mathbf{h}_{b,1}^{\mathrm{eff}}[n]\|^2\leq\|\mathbf{h}_{b,2}^{\mathrm{eff}}[n]\|^2\leq\cdots\leq\|\mathbf{h}_{b,K}^{\mathrm{eff}}[n]\|^2.
\end{align}
Under fixed successive interference cancellation (SIC) decoding order, terrestrial user $k$ successively decodes and cancels the signals of users $i\!<\!k$, while treating the signals of users $i\!>\!k$ as interference when decoding its own signal \cite{9316920}. An analogous fixed SIC decoding order is assumed for the satellite users.\\
After canceling the signals of users $i<k$, the residual received signal at terrestrial user $k$ is given by
\begin{align}
    \tilde{y}_{_k}[n] \!&=\! \mathbf{h}_{b,k}^{\mathrm{eff}}[n]\mathbf{v}_{b,k}[n]s_{b,k}[n] +\!\sum\nolimits_{i= k+1}^{K}\mathbf{h}_{b,k}^{\mathrm{eff}}[n]\mathbf{v}_{b,i}[n]s_{b,i}[n] \nonumber\\
&+\!\!\sum_{l=1}^{L}\!\mathbf{h}_{s,k}^{\mathrm{eff}}[n]\mathbf{v}_{s,l}[n]s_{s,l}[n]\!\!+\!\!\mathbf{h}_{U,k}^H[n]\boldsymbol{\Theta}_U[n]\mathbf{n}_U\!\!+\!n_k[n].
\end{align}
Similarly, after canceling the signals of users $j<l$, the residual received signal at satellite user $l$ is expressed as
\begin{align}
    \tilde{y}_{_l}[n] &= \mathbf{h}_{s,l}^{\mathrm{eff}}[n]\mathbf{v}_{s,l}[n]s_{s,l}[n] +\!\sum\nolimits_{j= l+1}^{L}\mathbf{h}_{s,l}^{\mathrm{eff}}[n]\mathbf{v}_{s,j}[n]s_{s,j}[n] \nonumber\\
&+\!\!\sum_{k=1}^{K}\!\!\mathbf{h}_{b,l}^{\mathrm{eff}}[n]\mathbf{v}_{b,k}[n]s_{b,k}[n]\!+\!\mathbf{h}_{H,l}^H[n]\boldsymbol{\Theta}_H[n]\mathbf{n}_H \!\!+\!\!n_l[n].
\end{align}
\subsection{SINR and Rate expressions}
Under a fixed SIC decoding order, the terrestrial user $k$ decodes the signals of the users $i\!<k$, while treating the signals of the users $i\!>k$ as residual intra-cell interference, along with cross-tier interference from the SAT. The corresponding SINRs for decoding user $i$'s signal ($i\!<\!k$) and the desired signal at terrestrial user $k$ are given in \eqref{TBS_SINR_i} and \eqref{TBS_SINR_k}, respectively.
Similarly, the satellite user $l$ decodes the signals of the users $j\!<\!l$ under a fixed SIC order, while treating the signals of the users $j\!>l$ as residual intra-system interference, along with cross-tier interference from the TBS. The corresponding SINRs for decoding user $j$'s signal ($j\!<l$) and the desired signal at user $l$ are given in \eqref{SAT_SINR_j} and \eqref{SAT_SINR_l}, respectively.
\par The achievable rates of terrestrial user $k$ and satellite user $l$ at time slot $n$ are given by
\begin{align}
    R_k[n] = \log_2(1+\gamma_k[n]),\quad R_l[n] = \log_2(1+\gamma_{l}[n]).
\end{align}
The corresponding average sum-rate over a transmission frame of $N$ time slots is expressed as
\begin{align}
    \mathcal{R} = \tfrac{1}{N}\sum\nolimits_{n=1}^N\Big(\sum\nolimits_{k=1}^KR_k[n]+\sum\nolimits_{l=1}^LR_l[n]\Big).
\end{align}
\section{Problem formulation}\label{ref:Problem formulation}
The objective of this work is to maximize the average sum-rate by jointly optimizing the transmit beamforming vectors, ARIS coefficients, and the 3D trajectories of the UAV and HAP over a transmission frame of $N$ time slots. Let $\mathcal{V}\!\!\triangleq\!\!\{\mathbf{v}_{b,k}[n],\mathbf{v}_{s,l}[n]\},\{\rho_{_{U,r}}[n],\theta_{U,r}[n]\},\{\rho_{_{H,\acute{r}}}[n],\theta_{H,\acute{r}}[n]\},\{\mathbf{q}_{_U}[n],\\\mathbf{q}_{_H}[n]\}$ denote the set of all optimization variables. Accordingly, the average sum-rate maximization problem is formulated as 
\begin{subequations} \label{OPT}
\begin{align}
\mathbf{(P1)}\quad&\max_{\mathcal{V}}\;\; 
 \frac{1}{N}\!\sum\nolimits_{n=1}^N\!\Big(\sum\nolimits_{k=1}^K\!R_k[n]\!+\!\sum\nolimits_{l=1}^L\!R_l[n]\!\Big) \label{opt:obj} \\
& \quad \text{s.t.} \;\;\; 
 \sum\nolimits_{k=1}^{K} \|\mathbf{v}_{b,k}[n]\|^2 \le P_b, \quad \forall\; n, \label{opt:TBS_beamforming} \\
& \qquad \quad \sum\nolimits_{l=1}^{L} \|\mathbf{v}_{s,l}[n]\|^2 \le P_s, \quad \forall \;n, \label{opt:SAT_beamforming} \\
&\qquad \quad  0 \le \rho_{_{U,r}}[n] \le \rho_{_U}^{max}, \quad \forall\;r,n, \label{opt:UAV_ARIS amplitude}\\
& \qquad \quad 0 \le \rho_{_{H,\acute{r}}}[n] \le \rho_{_H}^{max}, \quad \forall\;\acute{r},n, \label{opt:HAP_ARIS amplitude}\\
& \qquad \quad \sum\nolimits_{k=1}^K\big\|\boldsymbol{\Theta}_U[n]\mathbf{H}_{b,U}[n]\mathbf{v}_{b,k}[n]\big\|^2\nonumber\\&\hspace{6em}+\sigma_U^2\|\boldsymbol{\Theta}_U[n]\|^2_F\le P_U,\,\forall\;n, \label{opt:UAV_ARIS power}\\
& \qquad \quad \sum\nolimits_{l=1}^L\big\|\boldsymbol{\Theta}_H[n]\mathbf{H}_{s,H}[n]\mathbf{v}_{s,l}[n]\big\|^2\nonumber\\&\hspace{6em}+\sigma_H^2\|\boldsymbol{\Theta}_H[n]\|^2_F\le P_H,\,\forall\;n, \label{opt:HAP_ARIS power}\\
& \qquad \quad  \theta_{U,r}[n],\,\theta_{H,\acute{r}}[n]\in[0,2\pi),\quad \forall\; r,\acute{r},n, \label{opt:ARIS phase} \\
& \qquad \quad \|\mathbf{q}_{_U}[n+1]-\mathbf{q}_{_U}[n]\| \leq V_{U,max}\cdot\delta \;\; \forall\; n, \label{opt:UAV trajectories}\\
& \qquad \quad \|\mathbf{q}_{_H}[n+1]-\mathbf{q}_{_H}[n]\| \leq V_{H,max}\cdot \delta \;\; \forall\; n, \label{opt:HAP trajectories} \\
& \qquad \quad z_{_U}^{min} \le z_{_U}[n] \le z_{_U}^{max} \quad \forall\; n, \label{opt:UAV altitude} \\
& \qquad \quad z_{_H}^{min} \le z_{_H}[n] \le z_{_H}^{max} \quad \forall\; n, \label{opt:HAP altitude} \\
& \qquad \quad \mathbf{q}_{_U}[1] = \mathbf{q}_{_U}^{\text{init}}, \quad \mathbf{q}_{_U}[N] = \mathbf{q}_{_U}^{\text{final}}, \label{opt:UAV init-final} \\
& \qquad \quad  \mathbf{q}_{_H}[1] = \mathbf{q}_{_H}^{\text{init}}, \quad \mathbf{q}_{_H}[N] = \mathbf{q}_{_H}^{\text{final}}, \label{opt:HAP init-final}
\end{align}
\end{subequations}
where $P_b$ and $P_s$ denote the maximum transmit power budgets of TBS and SAT, respectively, while $P_U$ and $P_H$ represent the maximum amplification power budgets of UAV-ARIS and HAP-ARIS. The parameters $V_{U,max}$ and $V_{H,max}$ denote the maximum flying speeds of UAV and HAP, respectively. 
\par Constraints \eqref{opt:TBS_beamforming} and \eqref{opt:SAT_beamforming} limit the transmit power at TBS and SAT, respectively. Constraints \eqref{opt:UAV_ARIS amplitude} and \eqref{opt:HAP_ARIS amplitude} bound the amplification factors of the ARIS elements, while \eqref{opt:UAV_ARIS power} and \eqref{opt:HAP_ARIS power} limit the total amplification power at the UAV-ARIS and HAP-ARIS, respectively. The constraint \eqref{opt:ARIS phase} specifies the feasible phase shift range of the ARIS elements. Constraints (\ref{opt:UAV trajectories}) and (\ref{opt:HAP trajectories}) impose mobility limits, whereas \eqref{opt:UAV altitude} and \eqref{opt:HAP altitude} ensure feasible altitude ranges for UAV and HAP. Finally, \eqref{opt:UAV init-final} and \eqref{opt:HAP init-final} enforce the initial and final position constraints. 
\section{Proposed solution}\label{ref:Proposed solution}
Problem~$\mathbf{(P1)}$ is highly non-convex due to the coupling among optimization variables, the non-linear SINR expressions, ARIS amplification, and trajectory-dependent channels. Therefore, obtaining the global optimum is intractable. To address this challenge, we adopt a block coordinate descent (BCD) framework, in which the original problem is decomposed into four tractable subproblems. Specifically, each block of variables is optimized iteratively while keeping the remaining variables fixed. The proposed BCD algorithm is guaranteed to converge to a stationary point of $\mathbf{(P1)}$.
\vspace{-0.75em}
\subsection{Transmit Beamforming Optimization using WMMSE}
Given fixed UAV-ARIS and HAP-ARIS coefficients, as well as fixed UAV and HAP trajectories, the transmit beamforming vectors $\{\mathbf{v}_{b,k}[n],\mathbf{v}_{s,l}[n]\}$ at TBS and SAT are optimized. To this end, the average sum-rate maximization problem is transformed into an equivalent WMMSE subproblem by introducing auxiliary receive equalizers and mean-square error (MSE) weights. The resulting problem is then solved iteratively, where each variable is updated in closed-form using Lagrange multiplier methods.
\subsubsection{MSE Formulation}
For terrestrial user $k$, a linear receive equalizer $u_k[n] \in \mathbb{C}$ is introduced to estimate the desired signal $s_{b,k}[n]$ from the post-SIC received signal $\tilde{y}_k[n]$. The corresponding interference-plus-noise term, denoted by $\mathrm{T}_k$, and the MSE are defined as
\begin{align}
    \mathrm{T}_k[n] &\!=\! \sum\nolimits_{m=k}^K\!\big|\mathbf{h}_{b,k}^{\mathrm{eff}}[n]\mathbf{v}_{b,m}[n]\big|^2 \!+\! \sum\nolimits_{l=1}^L\! \big|\mathbf{h}_{s,k}^{\mathrm{eff}}[n]\mathbf{v}_{s,l}[n]\big|^2 \nonumber \\
    & \hspace{8em} + \big\|\mathbf{h}_{U,k}^H[n]\boldsymbol{\Theta}_U[n]\big\|^2\sigma_U^2 + \sigma_k^2, \\
    e_k[n] &= \big|u_k[n]\big|^2\mathrm{T}_k[n]-2\Re\left\{u_k[n]\mathbf{h}_{b,k}^{\mathrm{eff}}[n]\mathbf{v}_{b,k}[n]\right\}+1.
\end{align}
\par Similarly, for satellite user $l$, the interference-plus-noise term and the corresponding MSE after SIC are defined as
\begin{align}
    \mathrm{T}_l[n] &\!=\! \sum\nolimits_{m=l}^L\!\big|\mathbf{h}_{s,l}^{\mathrm{eff}}[n]\mathbf{v}_{s,m}[n]\big|^2 \!+\! \sum\nolimits_{k=1}^K\! \big|\mathbf{h}_{b,l}^{\mathrm{eff}}[n]\mathbf{v}_{b,k}[n]\big|^2 \nonumber \\
    & \hspace{8em} + \big\|\mathbf{h}_{H,l}^H[n]\boldsymbol{\Theta}_H[n]\big\|^2\sigma_H^2 + \sigma_l^2,\\
    e_l[n] &= \big|u_l[n]\big|^2\mathrm{T}_l[n]-2\Re\left\{u_l[n]\mathbf{h}_{s,l}^{\mathrm{eff}}[n]\mathbf{v}_{s,l}[n]\right\}+1.
\end{align}
\subsubsection{WMMSE Reformulation}
By introducing positive weights $w_k[n]$ and $w_l[n]$ for terrestrial and satellite users, respectively, the sum-rate maximization problem is equivalently reformulated as the following weighted sum-MSE minimization problem
\begin{align}
    \mathbf{(P2)}&\,\,\min_{\substack{\{\mathbf{v}_{b,k}[n],\mathbf{v}_{s.l}[n]\},\\\{u_k[n],u_l[n]\},\\\{w_k[n],w_l[n]\}}}\!\frac{1}{N}\!\!\sum_{n=1}^N\!\!\Big(\!\sum\nolimits_{k=1}^K\!\check{R}_k[n]\!+\!\!\sum\nolimits_{l=1}^L\!\check{R}_l[n]\!\Big)\\
&\qquad \qquad\text{s.t.} \quad 
\eqref{opt:TBS_beamforming} \;\&\; \eqref{opt:SAT_beamforming},
\end{align}
where 
\begin{align}
\check{R}_k[n] &= w_k[n]e_k[n]-\log w_k[n],\\ \check{R}_l[n] &= w_l[n]e_l[n]-\log w_l[n].
\end{align}
\subsubsection{Iterative WMMSE Updates}
The problem~$\mathbf{(P2)}$ is solved iteratively by alternately updating the receive equalizers $\{u_k[n],u_l[n]\}$, the MSE weights $\{w_k[n],w_l[n]\}$ and the transmit beamforming vectors $\{\mathbf{v}_{b,k}[n],\mathbf{v}_{s,l}[n]\}$.
\par \textit{a) Receiver update:}  
For given beamforming vectors, the optimal MMSE receive equalizers are expressed as
\begin{align}
u_k^{\star}[n] = \frac{\mathbf{h}_{b,k}^{\mathrm{eff}}[n]\mathbf{v}_{b,k}[n]}{\mathrm{T}_k[n]},\quad u_l^{\star}[n] = \frac{\mathbf{h}_{s,l}^{\mathrm{eff}}[n]\mathbf{v}_{s,l}[n]}{\mathrm{T}_l[n]}.
\end{align}
\par \textit{b) Weight update:}  
For given equalizers, the optimal MSE weights are given by
\begin{align}
w_k^{\star}[n] = \frac{1}{e_k[n]}, \quad w_l^{\star}[n] = \frac{1}{e_l[n]}.
\end{align}
\par \textit{c) Beamforming update:}  
With fixed $\{u_k[n],u_l[n]\}$ and $\{w_k[n],w_l[n]\}$, the optimal beamforming vectors at the TBS are given by
\begin{align}
    \mathbf{v}_{b,k}^{\star}[n] = \big(\mathbf{A}_b[n] + \lambda_b \mathbf{I}_{M_b}\big)^{-1}\mathbf{b}_{b,k}[n],
\end{align}
where $\lambda_b \geq 0$ is the Lagrangian multiplier selected via bisection to satisfy $\sum\nolimits_{k=1}^K\|\mathbf{v}_{b,k}[n]\|^2 = P_b$.\\
The matrix \!$\mathbf{A}_b\!\in\! \mathbb{C}^{M_b\!\times\! M_b}$\! and the vector \!$\mathbf{b}_{b,k}\!\in\! \mathbb{C}^{M_b}$\! are given by
\begin{align}
    \mathbf{A}_b[n] &= \sum\nolimits_{i=1}^K w_{i}[n]\big|u_{i}[n]\big|^2\big(\mathbf{h}_{b,i}^{\mathrm{eff}}[n]\big)^H\mathbf{h}_{b,i}^{\mathrm{eff}}[n]\nonumber\\ &\qquad + \sum\nolimits_{l=1}^Lw_l[n]\big|u_l[n]\big|^2\big(\mathbf{h}_{b,l}^{\mathrm{eff}}[n]\big)^H\mathbf{h}_{b,l}^{\mathrm{eff}}[n],\\
    \mathbf{b}_{b,k}[n] &= w_{k}[n]u_{k}^*[n]\big(\mathbf{h}_{b,k}^{\mathrm{eff}}[n]\big)^H.
\end{align}
\par Similarly, the optimal beamforming vectors at the SAT are given by
\begin{align}
    \mathbf{v}_{s,l}^{\star}[n] = \big(\mathbf{A}_s[n] + \lambda_s \mathbf{I}_{M_s}\big)^{-1}\mathbf{b}_{s,l}[n],
\end{align}
where $\lambda_s \ge 0$ is selected to satisfy $\sum\nolimits_{l=1}^L\|\mathbf{v}_{s,l}[n]\|^2 = P_s$.\\
The matrix \!$\mathbf{A}_s\!\in\! \mathbb{C}^{M_s\!\times\! M_s}$\! and the vector \!$\mathbf{b}_{s,l}\!\in\! \mathbb{C}^{M_s}$\! are given by
\begin{align}
    \mathbf{A}_s[n] &= \sum\nolimits_{j=1}^L w_{j}[n]\big|u_{j}[n]\big|^2\big(\mathbf{h}_{s,j}^{\mathrm{eff}}[n]\big)^H\mathbf{h}_{s,j}^{\mathrm{eff}}[n]\nonumber\\ &\quad \quad+ \sum\nolimits_{k=1}^Kw_k[n]\big|u_k[n]\big|^2\big(\mathbf{h}_{s,k}^{\mathrm{eff}}[n]\big)^H\mathbf{h}_{s,k}^{\mathrm{eff}}[n],\\
    \mathbf{b}_{s,l}[n] &= w_{l}[n]u_{l}^*[n]\big(\mathbf{h}_{s,l}^{\mathrm{eff}}[n]\big)^H.
\end{align}
The above WMMSE algorithm monotonically decreases the weighted sum-MSE objective and converges to a stationary point of problem~$\mathbf{(P2)}$, with all variables updated sequentially in each iteration.
\vspace{-0.75em}
\subsection{UAV-ARIS Optimization via Manifold Optimization and SCA}
In this subsection, the UAV-ARIS coefficients 
$\boldsymbol{\Theta}_U$ for a given time slot, while the beamforming vectors $\{\mathbf{v}_{b,k}, \mathbf{v}_{s,l}\}$, the WMMSE auxiliary variables $\{u_k, u_l, w_k, w_l\}$, and the UAV trajectory $\mathbf{q}_{_U}$ are fixed. For notational simplicity, the time-slot index 
$n$ is omitted in the following derivations. 
\par We define the UAV-ARIS coefficient vector as 
\begin{align}
    \boldsymbol{\varphi}_U \triangleq \left[\sqrt{\rho_{_{U,1}}}e^{\iota\theta_{U,1}},\cdots,\sqrt{\rho_{_{U,N_U}}}e^{\iota\theta_{U,N_U}}\right]^T,
\end{align}
such that $\boldsymbol{\Theta}_U=\mathrm{diag}(\boldsymbol{\varphi}_U)$. 
\par For terrestrial user $k$, we have
\begin{align}
    \mathbf{h}^H_{U,k}\boldsymbol{\Theta}_U\!=\! \boldsymbol{\varphi}_{U}^H\mathrm{diag}(\mathbf{h}_{U,k}),\;\;
    \mathbf{g}_{k,m}\!=\! \mathrm{diag}(\mathbf{h}_{U,k}\!)\mathbf{H}_{b,U}\mathbf{v}_{b,m}.
\end{align}
\par Similarly, for satellite user $l$, the TBS-to-user $l$ interference components dependent on the UAV-ARIS are expressed as
\begin{align}
    \mathbf{h}^H_{U,l}\boldsymbol{\Theta}_U\!=\! \boldsymbol{\varphi}_{U}^H\mathrm{diag}(\mathbf{h}_{U,l}),\quad
    \mathbf{g}_{l,k}\!=\! \mathrm{diag}(\mathbf{h}_{U,l})\mathbf{H}_{b,U}\mathbf{v}_{b,k}.
\end{align}
Based on the above definitions, the UAV-ARIS optimization problem at time slot $n$ is formulated as
\begin{subequations}
    \begin{align}
\mathbf{(P3)}\quad&\min_{\boldsymbol{\varphi}_{U}}\quad f(\boldsymbol{\varphi}_{U})=\boldsymbol{\varphi}_{U}^H\mathcal{Q}_U\boldsymbol{\varphi}_{U}-2\Re\big\{\boldsymbol{\varphi}_{U}^H\mathbf{\mathfrak{q}}_{_U}\big\}\\
&\quad\text{s.t.} \quad 0\le \big|[\boldsymbol{\varphi}_{U}]_{r}\big|^2\le \rho_{_U}^{max}, \quad \forall\,r,\\
& \qquad \quad \boldsymbol{\varphi}_{U}^H\Upsilon_U\boldsymbol{\varphi}_{U}\le P_U, 
%& \qquad \quad |\theta_{U,r}|=1,\quad \forall\,r,
\end{align}
\end{subequations}
where 
\begin{align}
    \mathcal{Q}_U &= \sum\nolimits_{k=1}^K\!\!w_k|u_k|^2\!\Big(\!\sum\nolimits_{m=k}^K\mathbf{g}_{k,m}\mathbf{g}_{k,m}^H\!+\!\sigma_U^2\mathrm{diag}(|\mathbf{h}_{U,k}|^2)\!\Big)\!\nonumber\\&\qquad\qquad+\sum\nolimits_{l=1}^Lw_l|u_l|^2\sum\nolimits_{k=1}^K\mathbf{g}_{l,k}\mathbf{g}_{l,k}^H,\\
    \mathbf{\mathfrak{q}}_{_U}&=\sum\nolimits_{k=1}^Kw_ku_k^*\mathrm{diag}(\mathbf{h}_{U,k})\mathbf{H}_{b,U}\mathbf{v}_{b,k},\\
    \Upsilon_U&=\sum\nolimits_{k=1}^K\mathrm{diag}(|\mathbf{H}_{b,U}\mathbf{v}_{b,k}|^2)+\sigma_U^2\mathbf{I}_{N_U}.
\end{align}
\par The problem~$\mathbf{(P3)}$ is nonconvex due to the coupling between the amplification factors and phase shifts in $\boldsymbol{\varphi}_U$, as well as the quadratic power constraint. To address this issue, we decompose $\boldsymbol{\varphi}_U$ into amplification factor and phase components and solve $\mathbf{(P3)}$ via alternating optimization. 
\par Specifically, $\boldsymbol{\varphi}_U$ is expressed as 
\begin{align}
    \boldsymbol{\varphi}_U = \mathbf{D}_U^{1/2}\boldsymbol{\theta}_U,
\end{align}
where $\mathbf{D}_U = \mathrm{diag}(\rho_{_{U,1}},\cdots,\rho_{_{U,N_U}})$
denotes the amplification factors matrix, and $\boldsymbol{\theta}_U=\left[e^{\iota\theta_{U,1}},\cdots,e^{\iota\theta_{U,N_U}}\right]^T$ denotes the UAV-ARIS phase shift vector that satisfies the unit-modulus constraint $|[\boldsymbol{\theta}_{U}]_{r}|=1,\quad \forall\,r$.
\subsubsection{UAV-ARIS Phase Optimization via Manifold Optimization}
With fixed $\mathbf{D}_U$, the problem~$\mathbf{(P3)}$ reduces to a phase-only optimization problem over a complex unit-modulus manifold. The resulting subproblem is formulated as
\begin{subequations}\label{UAV-ARIS_phase}
\begin{align}
\mathbf{(\!P3{-\!}1\!)}\;
&\min_{\boldsymbol{\theta}_U}f(\!\boldsymbol{\theta}_U\!)\!=\!
\boldsymbol{\theta}_U^H\! \mathbf{D}_U^{1/2}\!\mathcal{Q}_U\!\mathbf{D}_U^{1/2}\!\boldsymbol{\theta}_U\!\!-\!2\Re\!\big\{\!\boldsymbol{\theta}_U^H\!\mathbf{D}_U^{1/2}\!\!\mathbf{\mathfrak{q}}_{_U}\!\big\}\! \\
&\quad\text{s.t.}\quad \big|[\boldsymbol{\theta}_U]_{r}\big|=1,\quad \forall \;r.
\end{align}
\end{subequations}
The subproblem~$\mathbf{(P3\!-\!1)}$ remains non-convex due to unit-modulus constraints. To address this, we adopt a manifold-based RCG method over the complex unit-modulus manifold. 
Specifically, define the manifold as
\begin{align}\label{Manifold}
\mathcal{M} \triangleq
\big\{
\boldsymbol{\theta}_U\in\mathbb{C}^{N_U} :
\big|[\boldsymbol{\theta}_U]_{r}\big|=1,\ \forall \,r
\big\}.
\end{align}
The tangent space at $\boldsymbol{\theta}_U^{(t)}\in\mathcal{M}$ is given by
\begin{align}\label{Tangent}
\mathcal{T}_{\boldsymbol{\theta}_U^{(t)}}\mathcal{M}
=
\big\{
\boldsymbol{\eta}\in\mathbb{C}^{N_U} :
\Re\{\boldsymbol{\eta}\circ(\boldsymbol{\theta}_U^{(t)})^*\}=0
\big\}.
\end{align}
%which enforces orthogonality to the radial direction at each manifold point. 
Since $\mathcal{Q}_U$ is Hermitian positive semidefinite, the Euclidean gradient is given by
\begin{align}\label{Euclidean space}
\nabla f(\boldsymbol{\theta}_U) = 2\mathbf{D}_U^{1/2}\mathcal{Q}_U\mathbf{D}_U^{1/2}\boldsymbol{\theta}_U - 2\mathbf{D}_U^{1/2}\mathbf{\mathfrak{q}}_{_U},
\end{align}
and the Riemannian gradient is obtained via orthogonal projection onto the tangent space as
\begin{align}\label{Riemmanian gradient}
\mathrm{grad}\, f(\boldsymbol{\theta}_U) \!=\!
\nabla f(\boldsymbol{\theta}_U^{(t)})
\!-\!
\Re\!\left\{\!
\nabla f(\boldsymbol{\theta}_U^{(t)}) \circ (\boldsymbol{\theta}_U^{(t)})^*
\!\right\}
\!\circ \boldsymbol{\theta}_U^{(t)}.
\end{align}
The Riemannian conjugate gradient update is given by
\begin{align}\label{RCG_Update}
\boldsymbol{\zeta}^{(t+1)}=
-\mathrm{grad}\, f(\boldsymbol{\theta}_U^{(t)})
+
\chi^{(t+1)}
\mathsf{T}_{\boldsymbol{\theta}_U^{(t)}\to \boldsymbol{\theta}_U^{(t+1)}}(\boldsymbol{\zeta}^{(t)}),
\end{align}
where $\chi^{(t+1)}$ is the Polak-Ribière parameter \cite[Ch. ~8.3]{AbsilMahonySepulchre+2008}, and
$\mathsf{T}_{\boldsymbol{\theta}_U^{(t)}\to \boldsymbol{\theta}_U^{(t+1)}}(\cdot)$ denotes the vector transport, defined as
\begin{align}\label{Transport}
\mathsf{T}_{\boldsymbol{\theta}_U^{(t)}\to \boldsymbol{\theta}_U^{(t+1)}}(\boldsymbol{\zeta})
=
\boldsymbol{\zeta}
-
\Re\{\boldsymbol{\zeta}\circ(\boldsymbol{\theta}_U^{(t+1)})^*\}
\circ \boldsymbol{\theta}_U^{(t+1)}.
\end{align}
To ensure feasibility, a retraction is applied as 
\begin{align}\label{Retraction}
\mathsf{R}_{\boldsymbol{\theta}_U^{(t)}}(\xi^{(t)}\boldsymbol{\zeta}^{(t)})
=
\mathrm{unt}\big(\boldsymbol{\theta}_U^{(t)}+\xi^{(t)}\boldsymbol{\zeta}^{(t)}\big),
\end{align}
where $\mathrm{unt}(\cdot)$ denotes element-wise normalization, i.e., $[\mathrm{unt}(\mathbf{x})]_r \!=\! x_r/|x_r|$, and $\xi^{(t)}$ is determined via Armijo backtracking line search \cite[Ch. ~4.2]{AbsilMahonySepulchre+2008}.
\subsubsection{UAV-ARIS Amplification factor Optimization via SCA}
With fixed $\boldsymbol{\theta}_U$, the problem~$\mathbf{(P3)}$ reduces to optimizing the amplification factor $\boldsymbol{\rho}_{_U}$.
The UAV-ARIS amplification factor optimization subproblem is given by
\begin{subequations}\label{UAV-ARIS_amplitude}
\begin{align}
\mathbf{(\!P3{\!-\!}2\!)}\; 
&\min_{\boldsymbol{\rho}_{_U}} 
f(\boldsymbol{\rho}_U\!)\!=\!\boldsymbol{\theta}_U^H\! \mathbf{D}_U^{1/2}\!\mathcal{Q}_U\!\mathbf{D}_U^{1/2}\!\boldsymbol{\theta}_U\!\!-\!\!2\Re\!\big\{\boldsymbol{\theta}_U^H\!\mathbf{D}_U^{1/2}\!\!\mathbf{\mathfrak{q}}_{_U}\!\!\big\} \\
&\quad\text{s.t.}\quad 0\le\rho_{_{U,r}}\le\rho_{_U}^{max},\quad \forall r,\\
&\qquad \quad g(\boldsymbol{\rho}_{_U})=\boldsymbol{\theta}_U^H \mathbf{D}_U^{1/2}\Upsilon_U\mathbf{D}_U^{1/2}\boldsymbol{\theta}_U\le P_U.
\end{align}
\end{subequations}
The subproblem~$\mathbf{(P3\!-\!2)}$ is nonconvex due to the square-root coupling in $\mathbf{D}_U^{1/2}$ and the quadratic UAV-ARIS power constraint. To address this, we employ SCA.\\
Define $\boldsymbol{\rho}_{_U}\!=\!\big[\rho_{_{U,1}},\cdots,\rho_{_{U,N_U}}\big]^T$ and $\boldsymbol{\vartheta}_U\!=\!\mathbf{D}_U^{1/2}\boldsymbol{\theta}_U\!=\!\mathrm{diag}(\boldsymbol{\theta}_U)\sqrt{\boldsymbol{\rho}_{_U}}$, then the objective function in \eqref{UAV-ARIS_amplitude} can be reformulated as
\begin{align}
f(\boldsymbol{\rho}_{_U})=\boldsymbol{\vartheta}^H\mathcal{Q}_U\boldsymbol{\vartheta}-2\Re\big\{\boldsymbol{\vartheta}^H\mathbf{\mathfrak{q}}_{_U}\big\},
\end{align}
At iteration $t$, a convex surrogate is constructed via first-order Taylor expansion,
\begin{subequations}
\begin{align}
\mathbf{(P3-2-SCA)}\quad
&\min_{\boldsymbol{\rho}_{_U}}\;
\tilde{f}(\boldsymbol{\rho}_{_U}) \\
&\quad\text{s.t.}\quad
0\le \rho_{_{U,r}}\le \rho_{_U}^{\max},\quad \forall\; r, \\
&\qquad
\tilde{g}(\boldsymbol{\rho}_{_U}) \le P_U ,
\end{align}
\end{subequations}
where
\begin{align}
\tilde f(\boldsymbol{\rho}_{_U}) = f(\boldsymbol{\rho}_{_U}^{(t)})+
\sum\nolimits_{r=1}^{N_U}\left[\nabla f(\boldsymbol{\rho}_{_U}^{(t)})\right]_{r}\!\!
(\rho_{_{U,r}}\!-\rho_{_{U,r}}^{(t)}),
\end{align}
with gradient
\begin{align}
    \left[\nabla f(\boldsymbol{\rho}_{_U}^{(t)})\right]_{r}=\frac{2}{\sqrt{\rho_{_{U,r}}^{(t)}}}\Re\left\{\theta^*_{U,r}\left(\big[\mathcal{Q}_U\boldsymbol{\vartheta}^{(t)}_U\big]_{r}-\mathfrak{q}_{_{U,r}}\right)\right\},
\end{align}
and
\begin{align}
\tilde g(\boldsymbol{\rho}_{_U})
&=
g(\boldsymbol{\rho}_{_U}^{(t)})
+
\sum\nolimits_{r=1}^{N_U}\left[\nabla g(\boldsymbol{\rho}_{_U}^{(t)})\right]_{r}\!\!
(\rho_{_{U,r}}\!-\rho_{_{U,r}}^{(t)}),
\end{align}
with 
\begin{align}
    \left[\nabla g(\boldsymbol{\rho}_{_U}^{(t)})\right]_{r}\!=\!\frac{2}{\sqrt{\rho_{_{U,r}}^{(t)}}}\Re\Big\{\theta^*_{U,r}\Big[\Upsilon_U\boldsymbol{\vartheta}_U^{(t)}\Big]_{r}\Big\}.
\end{align}
The convex subproblem~$\mathbf{(P3\!-\!2\!-\!SCA)}$ can be efficiently solved using standard convex solvers \cite{Boyd_Vandenberghe_2004}. The solution is used to update $\boldsymbol{\rho}_{_U}^{(t+1)}$, and the procedure is repeated until convergence. The proposed SCA ensures a non-increasing objective value and convergence to a stationary point of subproblem~$\mathbf{(P3\!-\!2)}$.
The overall UAV-ARIS optimization alternates between phase-shifts and amplification factors, as summarized in \textbf{Algorithm~1}.
\begin{algorithm}[t]
\caption{UAV-ARIS Coefficients Optimization via Manifold Optimization and SCA}
\label{alg:UAV-ARIS} 
\small
\textbf{Input:} $\mathbf{v}_{b,k}$, $\mathbf{v}_{s,l}$, $u_k$, $u_l$, $w_k$, $w_l$, $\mathbf{q}_{_U}$, $\mathcal{Q}_U$, $\mathbf{\mathfrak{q}}_{_U}$, $\Upsilon_U$, $\rho_{_U}^{\max}$, $P_U$, $\varepsilon^{\mathrm{out}},\varepsilon^{\mathcal{M}},\varepsilon^{\mathrm{SCA}}$, $t_{\max}^{\mathrm{out}},t_{\max}^{\mathcal{M}},t_{\max}^{\mathrm{SCA}}$. \\
\textbf{Initialize:} $\boldsymbol{\rho}_U^{(0)}$, $\boldsymbol{\theta}_U^{(0)}$, set $t\leftarrow 0$; $\boldsymbol{\zeta}^{(0)} = -\mathrm{grad}\; f(\mathbf{\boldsymbol{\theta}}_U^{(0)})$.
\begin{algorithmic}[1]
\STATE Decompose $\boldsymbol{\varphi}_U = \mathbf{D}_U^{1/2}\boldsymbol{\theta}_U$.
\REPEAT
\STATE $\mathbf{D}_U^{(t)} = \mathrm{diag}(\boldsymbol{\rho}_U^{(t)})$
\STATE \textbf{Phase optimization:} Solve $(\mathbf{P3-1})$ on manifold $\mathcal{M}$.
\STATE Initialize manifold iteration $i=0$.
\REPEAT
\STATE Compute the Euclidean gradient $\nabla f(\boldsymbol{\theta}_U^{(i)})$ using \eqref{Euclidean space}.
\STATE Obtain the Riemannian gradient by projecting onto the tangent space using \eqref{Riemmanian gradient}.
\STATE Update the conjugate search direction $\boldsymbol{\zeta}^{(i)}$ using the Polak-Ribière rule in \eqref{RCG_Update}.
\STATE Determine the step size $\xi^{(i)}$ via Armijo backtracking line search.
\STATE Update $\boldsymbol{\theta}_U^{(i+1)}$ using retraction \eqref{Retraction}.
\STATE $i \leftarrow i+1$.
\UNTIL{$\|\mathrm{grad}\,f(\boldsymbol{\theta}_U^{(i)})\| \le \varepsilon^{\mathcal{M}}$} or $i = t_{\max}^{\mathcal{M}}$
\STATE $\boldsymbol{\theta}_U^{(t+1)} \leftarrow \boldsymbol{\theta}_U^{(i)}$
\STATE \textbf{Amplification optimization:} Solve $(\mathbf{P3-2})$ via SCA.
\STATE Initialize SCA iteration $j=0$.
\REPEAT
\STATE Construct the first-order Taylor approximations $\tilde{f}(\boldsymbol{\rho}_U)$ and $\tilde{g}(\boldsymbol{\rho}_U)$ around $\boldsymbol{\rho}_U^{(j)}$.
\STATE Solve the convex problem $(\mathbf{P3-2-SCA})$.
\STATE Update the amplification vector $\boldsymbol{\rho}_U^{(j+1)}$.
\STATE $j \leftarrow j+1$.
\UNTIL{$|f(\boldsymbol{\rho}_U^{(j)})-f(\boldsymbol{\rho}_U^{(j-1)})| \le \varepsilon^{\mathrm{SCA}}$} or $j = t_{\max}^{\mathrm{SCA}}$
\STATE $\boldsymbol{\rho}_U^{(t+1)} \leftarrow \boldsymbol{\rho}_U^{(j)}$
\STATE $t \leftarrow t+1$.
\UNTIL $\big|f(\boldsymbol{\varphi}_U^{(t)}) -
f(\boldsymbol{\varphi}_U^{(t-1)})\big| \le \varepsilon^{\mathrm{out}}$ or $t = t_{\max}^{\mathrm{out}}$

\STATE $\boldsymbol{\varphi}_U^\star = \mathbf{D}_U^{1/2}\boldsymbol{\theta}_U$.
\end{algorithmic}
\textbf{Output:} Optimized UAV-ARIS coefficient vector $\boldsymbol{\varphi}_U^\star$.
\end{algorithm}
\subsection{HAP-ARIS Optimization via Manifold Optimization and SCA}
In this subsection, the HAP-ARIS coefficients $\boldsymbol{\Theta}_H$ are optimized for a given time slot, while the beamforming vectors $\{\mathbf{v}_{b,k},\mathbf{v}_{s,l}\}$, the WMMSE auxiliary variables $\{u_k,u_l,w_k,w_l\}$, and the HAP trajectory $\mathbf{q}_{_H}$ are fixed. Similar to the UAV-ARIS case, the resulting problem is nonconvex due to the unit-modulus constraints on the phase shifts and the quadratic ARIS power constraint.\\
We define the HAP-ARIS coefficient vector as
\begin{align}
\boldsymbol{\varphi}_H \triangleq 
\Big[\sqrt{\rho_{_{H,1}}}e^{\iota\theta_{H,1}},\cdots,\sqrt{\rho_{_{H,N_H}}}e^{\iota\theta_{H,N_H}}\Big]^T,
\end{align}
such that $\boldsymbol{\Theta}_H=\mathrm{diag}(\boldsymbol{\varphi}_H)$.
The HAP-ARIS optimization problem is formulated as
\begin{subequations}
\begin{align}
\mathbf{(P4)}\quad&
\min_{\boldsymbol{\varphi}_H}\quad
\boldsymbol{\varphi}_H^H\mathcal{Q}_H\boldsymbol{\varphi}_H
-2\Re\{\boldsymbol{\varphi}_H^H\mathfrak{q}_{_H}\} \\
&\quad\text{s.t.}\quad
0 \le |[\boldsymbol{\varphi}_H]_{r}|^2 \le \rho_{_H}^{\max},\quad \forall \;r,\\
&\qquad \quad
\boldsymbol{\varphi}_H^H\Upsilon_H\boldsymbol{\varphi}_H \le P_H,
\end{align}
\end{subequations}
where $\mathcal{Q}_H$, $\mathfrak{q}_{_H}$, and $\Upsilon_H$ are defined analogously to the UAV-ARIS case by replacing UAV-related channels with the corresponding HAP-related channels.\\
By decomposing $\boldsymbol{\varphi}_H=\mathbf{D}_H^{1/2}\boldsymbol{\theta}_H$, where $\mathbf{D}_H$ contains the HAP-ARIS amplification factors and $\boldsymbol{\theta}_H$ is the unit-modulus phase vector, the problem~$\mathbf{(P4)}$ can be solved via alternating optimization.
Specifically, the phase-shift vector $\boldsymbol{\theta}_H$ is optimized on a complex unit-modulus manifold using the RCG method, while amplification factors are optimized via SCA. Since the resulting subproblems share the same structure as those in the UAV-ARIS optimization subproblem, the detailed derivations are omitted for brevity.
\subsection{UAV and HAP Trajectory Optimization via SCA}
With the beamforming vectors and ARIS coefficient matrices obtained from the previous problems, the UAV and HAP trajectory optimization is formulated as
\begin{subequations} \label{Objective-Trajectory}
\begin{align}
\mathbf{(P5)}&\max_{\substack{ \{\mathbf{q}_{_U}[n],\mathbf{q}_{_H}[n]\}}} 
 \!\!\frac{1}{N}\!\!\sum\nolimits_{n=1}^{N}\!\! \Big(\! \sum\nolimits_{k=1}^{K}\!\! R_k[n] \!+ \!\!\!\sum\nolimits_{l=1}^{L} \!R_l[n]\! \Big) \\
&\qquad \text{s.t.} \quad 
\eqref{opt:UAV trajectories},\eqref{opt:HAP trajectories},\eqref{opt:UAV altitude},\eqref{opt:HAP altitude},\eqref{opt:UAV init-final},\eqref{opt:HAP init-final}. 
\end{align}
\end{subequations}
Problem~$\mathbf{(P5)}$ is non-convex due to the distance-dependent channel gains in the achievable rate expressions. To address this challenge, an SCA framework is adopted, where convex affine lower bounds are constructed for the trajectory-dependent received signal powers. 
Since the positions of TBS, SAT, and users are fixed, only the cascaded TBS-to-UAV-ARIS-to-user \!$k$ and SAT-to-HAP-ARIS-to-user \!$l$ links depend on the UAV and HAP trajectories, respectively. The direct TBS-to-user $k$ and SAT-to-user $l$ links are independent of trajectories and are thus omitted from trajectory optimization. 
\subsubsection{UAV Trajectory Surrogate Construction}
For terrestrial user $k$, the received signal power corresponding to the TBS-to-UAV-ARIS-to-user link is expressed as
\begin{align}
    \mathbf{p}(\mathbf{q}_{_U}) = C_U d_{b,U}^{-\eta_{_{b,U}}}d_{U,k}^{-\eta_{_{U,k}}}, 
\end{align}
where $C_U >0$ collects all distance-independent terms, including beamforming gains, UAV-ARIS coefficients, array responses, Rician fading components, and transmit power. These terms remain constant during each SCA iteration.\\
Define the auxiliary function,
\begin{align}
    h(d_{b,U},d_{U,k}) = d_{b,U}^{-\eta_{_{b,U}}}d_{U,k}^{-\eta_{_{U,k}}}.
\end{align}
Since $d^{-\eta}$ is convex for $d>0$ and $\eta>0$, the function $h(d_{b,U},d_{U,k})$ is convex with respect to $(d_{b,U},d_{U,k})$. Therefore, its first-order Taylor expansion gives a global affine lower bound.
Let $d_{b,U}^{(t)}$ and $d_{U,k}^{(t)}$ denote the distances at iteration $t$.
The gradients of $h(d_{b,U},d_{U,k})$ are given by
\begin{align}
\nabla_{d_{b,U}}h &= -\eta_{_{b,U}}d_{b,U}^{-\eta_{_{b,U}}-1}d_{U,k}^{-\eta_{_{U,k}}},\\ 
\nabla_{d_{U,k}}h &= -\eta_{_{U,k}}d_{b,U}^{-\eta_{_{b,U}}}d_{U,k}^{-\eta_{_{U,k}}-1}.
\end{align}
Applying the first-order Taylor expansion at $\!\big(\!d_{b,U}^{(t)},\!d_{U,k}^{(t)}\big)\!$ yields\cite{9656117}
\begin{align}
    h(d_{b,U},d_{U,k}) \geq h^{(t)} &+\nabla_{d_{b,U}}h^{(t)}\big(d_{b,U}-d_{b,U}^{(t)}\big)\nonumber\\&\quad+\nabla_{d_{U,k}}h^{(t)}\big(d_{U,k}-d_{U,k}^{(t)}\big),
\end{align}
where $h^{(t)}={d^{(t)}_{b,U}}^{-\eta_{_{b,U}}}{d^{(t)}_{U,k}}^{-\eta_{_{U,k}}}$.
\par Next substituting $d_{b,U}=\|\mathbf{q}_{_{TBS}}\!-\!\mathbf{q}_{_U}\|$ and  $d_{U,k} \!=\! \|\mathbf{q}_{_U}\!-\!\mathbf{q}_k\|$, the Euclidean norms are lower bounded via first-order approximations \cite{9656117}
\begin{subequations}
    \begin{align}
        \|\mathbf{q}_{_{TBS}}-\mathbf{q}_{_U}\| &\geq d_{b,U}^{(t)} + \tfrac{(\mathbf{q}_{_{TBS}}-\mathbf{q}_{_U}^{(t)})^T}{d_{b,U}^{(t)}}\big(\mathbf{q}_{_U}-\mathbf{q}_{_U}^{(t)}\big),\label{Eq:UAV Linear Norms1}\\
        \|\mathbf{q}_{_U}-\mathbf{q}_k\| &\geq d_{U,k}^{(t)} + \tfrac{(\mathbf{q}_{_U}^{(t)}-\mathbf{q}_{_k})^T}{d_{U,k}^{(t)}}\big(\mathbf{q}_{_U}-\mathbf{q}_{_U}^{(t)}\big).\label{Eq:UAV Linear Norms2}
    \end{align}
\end{subequations}
Substituting \eqref{Eq:UAV Linear Norms1} and \eqref{Eq:UAV Linear Norms2} into the Taylor expansion yields the following affine surrogate of the received power \cite{9656117},
\begin{align}
    \hat{\mathbf{p}}^{(t)}(\mathbf{q}_{_U}) = C_U \Big(h^{(t)}&+\nabla_{d_{b,U}}h^{(t)}\tfrac{(\mathbf{q}_{_{TBS}}-\mathbf{q}_{_U}^{(t)})^T}{d_{b,U}^{(t)}}\big(\mathbf{q}_{_U}-\mathbf{q}_{_U}^{(t)}\big)\nonumber\\&+\nabla_{d_{U,k}}h^{(t)}\tfrac{(\mathbf{q}_{_U}^{(t)}-\mathbf{q}_{_k})^T}{d_{U,k}^{(t)}}\big(\mathbf{q}_{_U}\!-\!\mathbf{q}_{_U}^{(t)}\big)\!\Big),
    \label{Eq:UAV Power Surrogate}
\end{align}
which is affine in $\mathbf{q}_{_U}$, and serves as a global lower bound of $\mathbf{p}(\mathbf{q}_{_U})$, with equality at $\mathbf q_{_U}=\mathbf q_{_U}^{(t)}$.
\subsubsection{HAP Trajectory Surrogate Construction}
For satellite user $l$, the received signal power corresponding to the SAT-to-HAP-ARIS-to-user $l$ link is given by
\begin{align}
\mathbf p(\mathbf q_{_H})=C_H d_{s,H}^{-2} d_{H,l}^{-2},
\end{align}
where $C_H>0$ collects all distance-independent terms, including satellite beamforming gains, HAP-ARIS coefficients, array responses, Rician fading components, and transmit power.\\
Following the same SCA procedure as in the UAV case, define
\begin{align}
\ddot{h}(d_{s,H},d_{H,l}) = d_{s,H}^{-2} d_{H,l}^{-2},
\end{align}
which is convex for $d_{s,H}>0$ and $d_{H,l}>0$.\\ Let $d_{s,H}^{(t)}$ and $d_{H,l}^{(t)}$ denote the distances at iteration $t$. The gradients of $\ddot{h}(d_{s,H},d_{H,l})$ are given by
\begin{align}
    \nabla_{d_{s,H}}\ddot{h} = -2d_{s,H}^{-3}d_{H,l}^{-2},\qquad \nabla_{d_{H,l}}\ddot{h} = -2d_{s,H}^{-2}d_{H,l}^{-3}.
\end{align}
Applying the first-order Taylor expansion at $\big(d_{s,H}^{(t)},d_{H,l}^{(t)}\big)$ yields
\begin{align}
\ddot{h}(d_{s,H},d_{H,l})
\geq
\ddot{h}^{(t)}
&+
\nabla_{d_{s,H}} \ddot{h}^{(t)}
\big(d_{s,H}-d_{s,H}^{(t)}\big)
\nonumber\\
&\quad+
\nabla_{d_{H,l}} \ddot{h}^{(t)}
\big(d_{H,l}-d_{H,l}^{(t)}\big),
\end{align}
where $\ddot{h}^{(t)}={d_{s,H}^{(t)}}^{-2}{d_{H,l}^{(t)}}^{-2}$.
\par Substituting $d_{s,H}\!=\!\|\mathbf q_{_{SAT}}\!\!-\!\mathbf q_{_H}\|$ and $d_{H,l}\!=\!\|\mathbf q_{_H}\!\!-\!\mathbf q_{_l}\|$, and applying first-order approximations of the Euclidean norms as in \eqref{Eq:UAV Linear Norms1}–\eqref{Eq:UAV Linear Norms2}, the affine surrogate of the received signal power is obtained as
\begin{align}
\hat{\mathbf p}^{(t)}(\mathbf q_{_H})
=
C_H \Big(
\ddot{h}^{(t)}
&+
\nabla_{d_{s,H}} \ddot{h}^{(t)}
\tfrac{(\mathbf q_{_{SAT}}-\mathbf q_{_H}^{(t)})^T}{d_{s,H}^{(t)}}
\big(\mathbf q_{_H}-\mathbf q_{_H}^{(t)}\big)
\nonumber\\
&+
\nabla_{d_{H,l}} \ddot{h}^{(t)}
\tfrac{(\mathbf q_{_H}^{(t)}-\mathbf q_{_l})^T}{d_{H,l}^{(t)}}
\big(\mathbf q_{_H}\!-\!\mathbf q_{_H}^{(t)}\big)\!
\Big),
\end{align}
which is affine in $\mathbf q_{_H}$ and serves as a global lower bound of $\mathbf p(\mathbf q_{_H})$, with equality at $\mathbf q_{_H}=\mathbf q_{_H}^{(t)}$.
\subsubsection{Surrogate SINR and Rate Expressions}
For each time slot $n$, the UAV and HAP trajectories $\big\{\mathbf{q}_{_U}[n],\mathbf{q}_{_H}[n]\big\}$ are optimized. Let $\mathcal{I}_k[n]$ and $\mathcal{I}_l[n]$ denote the interference-plus-noise power at terrestrial users $k$ and satellite users $l$, respectively. 
\par Using affine power surrogates, the corresponding surrogate SINRs are expressed as
\begin{align}
    \hat{\gamma}_k^{(t)}[n] = \frac{\hat{\mathbf{p}}^{(t)}(\mathbf{q}_{_U}[n])}{\mathcal{I}_k[n]},\quad \hat{\gamma}_l^{(t)}[n] = \frac{\hat{\mathbf{p}}^{(t)}(\mathbf{q}_{_H}[n])}{\mathcal{I}_l[n]}.
\end{align}
Accordingly, the surrogate rates are given by
\begin{align}
    \hat{R}_k[n] = \log_2(1+\hat{\gamma}_k^{(t)}[n]), \quad \hat{R}_l[n] = \log_2(1+\hat{\gamma}_l^{(t)}[n]).
\end{align}
The resulting convex subproblem is formulated as
\begin{subequations} \label{Surrogate-Objective-Trajectory}
\begin{align}
\mathbf{\!\!(P5\!-\!\!1\!)}\!\!\!&\max_{\substack{ \{\!\mathbf{q}_{_U}\![n],\mathbf{q}_{_H}\![n]\!\}}} 
 \!\!\tfrac{1}{N}\!\!\sum\nolimits_{n=1}^{N}\!\! \Big(\! \sum\nolimits_{k=1}^{K}\!\! \hat{R}_k[n] \!+ \!\!\sum\nolimits_{l=1}^{L} \!\hat{R}_l[n]\! \Big) \\
&\qquad \text{s.t.} \quad 
\eqref{opt:UAV trajectories},\eqref{opt:HAP trajectories},\eqref{opt:UAV altitude},\eqref{opt:HAP altitude},\eqref{opt:UAV init-final},\eqref{opt:HAP init-final}. 
\end{align}
\end{subequations}
\par Subproblem~$\mathbf{(P5\!-\!1)}$ is a convex QCQP, as the surrogates are affine in the trajectory variables and can be efficiently solved using standard convex optimization tools such as CVX \cite{Boyd_Vandenberghe_2004}.
\subsection{Overall Algorithm}
%%%%%%%%% Algorithm 2 %%%%%%%%%%%
\begin{algorithm}[t]
\caption{Proposed BCD-based Joint Beamforming, ARIS Coefficients, and Trajectory Optimization}
\label{alg:BCD} 
\small
\textbf{Input:} System parameters, channel statistics, power budgets $\{P_b,P_s\}$, tolerance $\varepsilon$,
maximum iterations $t_{\max}$. \\
\textbf{Initialize:} Beamforming vectors $\{\mathbf v_{b,k}^{(0)},\mathbf v_{s,l}^{(0)}\}$; UAV-ARIS and HAP-ARIS coefficient vectors $\{\boldsymbol{\varphi}_U^{(0)},\boldsymbol{\varphi}_{H}^{(0)}\}$;
UAV and HAP trajectories $\{\mathbf q_{_U}^{(0)},\mathbf q_{_H}^{(0)}\}$;
set outer iteration index $t \leftarrow 0$.
\begin{algorithmic}[1]
\REPEAT
\STATE \textbf{Beamforming update:} \textit{WMMSE} 
\STATE Update the WMMSE auxiliary variables $\{u_k,u_l\}$, weights $\{w_k,w_l\}$, and beamforming vectors $\{\mathbf v_{b,k}^{(t+1)},\mathbf v_{s,l}^{(t+1)}\}$ until convergence.

\STATE \textbf{UAV-ARIS optimization:} \textit{Manifold optimization} and \textit{SCA}
\STATE Update the UAV-ARIS coefficients 
$\{\boldsymbol{\theta}_U^{(t+1)},\boldsymbol{\rho}_U^{(t+1)}\}$
according to \textbf{Algorithm~1}.

\STATE \textbf{HAP-ARIS optimization:} \textit{Manifold optimization} and \textit{SCA}
\STATE Update the HAP-ARIS coefficients 
$\{\boldsymbol{\theta}_H^{(t+1)},\boldsymbol{\rho}_H^{(t+1)}\}$
using the same procedure as in \textbf{Algorithm~1}.

\STATE \textbf{Trajectory optimization:} \textit{SCA}
\STATE Update the UAV and HAP trajectories $\{\mathbf q_{_U}^{(t+1)},\mathbf q_{_H}^{(t+1)}\}$
by solving the convex trajectory subproblem $\mathbf{(P5\!-\!1)}$ via SCA.

\STATE $t \leftarrow t+1$.
\UNTIL $\big|\mathcal{R}^{(t)}-\mathcal{R}^{(t-1)}\big| \le \varepsilon$ or $t = t_{\max}$.

\end{algorithmic}
\textbf{Output:} Optimized  $\{\mathbf v_{b,k}^\star,\mathbf v_{s,l}^\star\}$,  $\{\boldsymbol{\varphi}_U^{\star},\boldsymbol{\varphi}_{H}^{\star}\}$, and  $\{\mathbf q_{_U}^\star,\mathbf q_{_H}^\star\}$.
\end{algorithm}
To jointly optimize the transmit beamforming vectors, ARIS coefficients, and the 3D trajectories of the UAV and HAP, we adopt a BCD framework \cite{11030800}, in which the original non-convex problem is decomposed into four tractable subproblems that are solved iteratively. Specifically, for given UAV-ARIS and HAP-ARIS coefficients as well as UAV and HAP trajectories, the transmit beamforming vectors are optimized using the WMMSE, which ensures a monotonic improvement of the average sum-rate and converges to a stationary point.
For given transmit beamforming vectors and trajectories, the UAV-ARIS and HAP-ARIS coefficient vectors are decomposed into phase-shift and amplification factor components. The phase-shift optimization is addressed using manifold-based RCG method, while amplification factors are optimized via SCA.
For given transmit beamforming vectors and ARIS coefficients, the UAV and HAP trajectories are optimized using the SCA, where a sequence of convex surrogate problems is solved iteratively. 
\par At the $t$-th BCD iteration, the effective channels are updated according to the initial ARIS coefficients and trajectories. Subsequently, the transmit beamforming vectors, UAV-ARIS and HAP-ARIS coefficients, and UAV/HAP trajectories are updated sequentially using WMMSE, manifold optimization combined with SCA, and SCA-based trajectory optimization, respectively. Each subproblem is solved to a prescribed accuracy, determined by convergence tolerances and maximum number of iterations. The BCD algorithm terminates when the increment of the average sum-rate between two consecutive iterations falls below a predefined threshold or when the maximum number of BCD iterations is reached. The overall procedure is summarized in \textbf{Algorithm~2}.
%%%%%%%% Simulation Parameters %%%%
\begin{table*}[t]
\centering
\caption{Simulation Parameters}
\label{tab:sim_param}
\begin{tabular}{|l|c|l|l|c|l|}%{|p{3.5cm}|p{1cm}|p{1.3cm}|p{3.5cm}|p{1cm}|p{1.3cm}|}
\hline
\multicolumn{1}{|c|}{\textbf{Parameter}} & \textbf{Symbol} & \multicolumn{1}{c|}{\textbf{Value}} & \multicolumn{1}{c|}{\textbf{Parameter}} & \textbf{Symbol} & \multicolumn{1}{c|}{\textbf{Value}} \\ \hline
Time slots  & $N$ & 60 & Slot duration  & $\delta$ & 1 s \\
TBS antennas  & $M_b$ & 8 & SAT antennas  & $M_s$ & 32 \\
Terrestrial users  & $K$ & 3 & Satellite users  & $L$ & 4 \\
UAV-ARIS elements  & $N_U$ & 16 & HAP-ARIS elements  & $N_H$ & 36 \\
UAV altitude range & $z_{_U}$ & [50,150] m & HAP altitude range & $z_{_H}$ & [18,22] km  \\ 
TBS carrier frequency  & $f_c$ & 3.5 GHz & SAT carrier frequency  & $\check{f}_c$ & 20 GHz \\ 
Reference path-loss  & $\beta_0$ & 0 dBm & Path-loss exponent & $\eta_{_{b,k}}$ & 3.5\\
Path-loss exponent & $\eta_{_{b,U}}$ & 2.2 & Path-loss exponent & $\eta_{_{U,k}}$ & 2.2 \\
SAT altitude  & $H_{SAT}$ & 600 km & SAT antenna gain  & $G_{SAT}$ & 30 dBi \cite{3GPP_TR_38_821}  \\
User antenna gain  & $G_l$ & 0 dBi \cite{3GPP_TR_38_901} & HAP antenna gain  & $G_{HAP}$ & 14 dBi \cite{3GPP_TR_38_821} \\
Rain attenuation mean & $\mu_r$ & -2.6 dB \cite{3GPP_TR_38_901} & Rain attenuation variance & $\sigma_r^2$ & 1.63 dB \cite{3GPP_TR_38_901} \\ 
TBS transmit power  & $P_b$ & 43 dBm & SAT transmit power  & $P_s$ & 54.77 dBm \\
TBS noise power & $\sigma_k^2$ & -90 dBm & SAT noise power &  $\sigma_l^2$ & -90 dBm \\
UAV-ARIS power budget & $P_U$ & 15 dBm & HAP-ARIS power budget & $P_H$ & 25 dBm \\
UAV-ARIS noise power & $\sigma_U^2$ & -84 dBm & HAP-ARIS noise power &  $\sigma_H^2$ & -113 dBm \\
Max. UAV speed  & $V_{U,max}$ & 30 m/s & Max. HAP speed & $V_{H,max}$ & 5 m/s \\ \hline
\multicolumn{6}{|c|}{Rician factors}  \\ \hline
TBS-to-terrestrial users & $\kappa_{_{b,k}}$ & 5 & SAT-to-satellite users & $\kappa_{_{s,l}}$ & 3\\
TBS-to-UAV-ARIS & $\kappa_{_{b,U}}$ & 8 & SAT-to-HAP-ARIS & $\kappa_{_{SAT,H}}$ & 6\\
UAV-ARIS-to-terrestrial users & $\kappa_{_{U,k}}$ & 5 & HAP-ARIS-to-satellite users & $\kappa_{_{H,l}}$ & 3\\ 
TBS-to-satellite users & $\kappa_{_{b,l}}$ & 5 & SAT-to-terrestrial users & $\kappa_{_{s,k}}$ & 3\\ 
UAV-ARIS-to-satellite users & $\kappa_{_{U,l}}$ & 5 & HAP-ARIS-to-terrestrial users & $\kappa_{_{H,k}}$ & 3\\  \hline
\end{tabular}
\vspace{-1.25em}
\end{table*}
%%%%%%%%%%%%%%%%%%%%%%%%%%%%%%%%%
\par \textit{Complexity Analysis:} The computational complexity of the proposed \textbf{Algorithm~2} comes mainly from four subproblems, namely transmit beamforming optimization at the TBS and SAT, UAV-ARIS optimization, HAP-ARIS optimization, and UAV/HAP trajectory optimization. The transmit beamforming optimization is solved using the WMMSE algorithm, whose complexity is dominated by matrix inversion, given by $\mathcal{O}\big(I_{\mathcal{W}}N(KM_b^3+LM_s^3)\big)$, where $I_{\mathcal{W}}$ denotes the number of WMMSE iterations. The UAV-ARIS coefficients are optimized using a manifold-based RCG method for RIS phase shifts and SCA for amplification factors. The corresponding complexity is  $\mathcal{O}\big(I_{\mathrm{out}}^UN(I_{\mathcal{M}_U}KN_U+I_{\rho_{_U}} N_U^3)\big)$, where $I_{\mathrm{out}}^U$, $I_{\mathcal{M}_U}$ and $I_{\rho_{_U}}$ denote the number of outer iterations of the UAV-ARIS subproblem, RCG iterations and SCA iterations, respectively. Similar the HAP-ARIS optimization has complexity $\mathcal{O}\big(I_{\mathrm{out}}^HN(I_{\mathcal{M}_H}LN_H+I_{\rho_{_H}} N_H^3)\big)$, where $I_{\mathrm{out}}^H$, $I_{\mathcal{M}_H}$ and $I_{\rho_{_H}}$ are number of outer iterations of the HAP-ARIS subproblem, RCG iterations and SCA iterations, respectively. The trajectory optimization for both UAV and HAP is solved using SCA with first-order Taylor approximations, resulting in a complexity of $\mathcal{O}(I_{\mathrm{traj}}N^2)$. Therefore, the overall computational complexity of the proposed algorithm is expressed as $\mathcal{O}\big(\mathcal{I}_{\mathrm{out}}N\big(I_{\mathcal{W}}(KM_b^3+LM_s^3)+I_{\mathrm{out}}^U(I_{\mathcal{M}_U}KN_U+I_{\rho_{_U}} N_U^3)+I_{\mathrm{out}}^H(I_{\mathcal{M}_H}LN_H+I_{\rho_{_H}} N_H^3)+I_{\mathrm{traj}}N^2\big)\big)$, where $\mathcal{I}_{\mathrm{out}}$ denotes the number of outer BCD iterations.
\section{Simulation results}\label{ref:Simulation results}
In this section, we present simulation results to evaluate the performance of the proposed BCD-based optimization framework and compare it with benchmark schemes. The TBS and SAT are located at fixed positions given by $\mathbf{q}_{_{TBS}} = (0,0,0)$ m and $\mathbf{q}_{_{SAT}} = (0.6,0.8,600)$ km, respectively. Terrestrial users are independently and uniformly distributed within a circular region of radius $300$ m on the ground plane, with the TBS located at the center. Meanwhile, satellite users are independently and uniformly distributed within a separate circular region of radius $500$ m centered at $[1500,800]$ m on the ground plane. All users are assumed to be located at ground level (zero altitude). The UAV operates within an altitude range of $[50,150]$ m. Its initial position is set directly above the TBS, whereas its final position is located above the centroid of the terrestrial users. The UAV trajectory is initialized as a straight-line path between the initial and final positions at a fixed altitude of $100$ m. The HAP operates within an altitude range of $[18,22]$ km and is deployed above the satellite user region. Its trajectory is initialized as a straight-line path at a fixed altitude of $20$ km, centered over the satellite users. Both UAV and HAP trajectories are discretized into $N$ time slots. The detailed simulation parameters are summarized in Table \ref{tab:sim_param}. We considered the following benchmarks: \textit{No RIS}, where the RIS is removed from the system, \textit{random RIS}, where the RIS is deployed, but its phase shifts are randomly generated and fixed without optimization, \textit{passive RIS}, where the RIS only adjusts the phase shifts without amplification, and Fixed trajectory, where the UAV moves along a predetermined straight-line path with a constant speed and without trajectory optimization.
%%%%%%%%%% Figure Section -1 %%%%%%%%%%
\begin{figure*}[t!]
\centering
\begin{subfigure}{0.335\textwidth} 
\centerline{\includegraphics[scale=0.43] {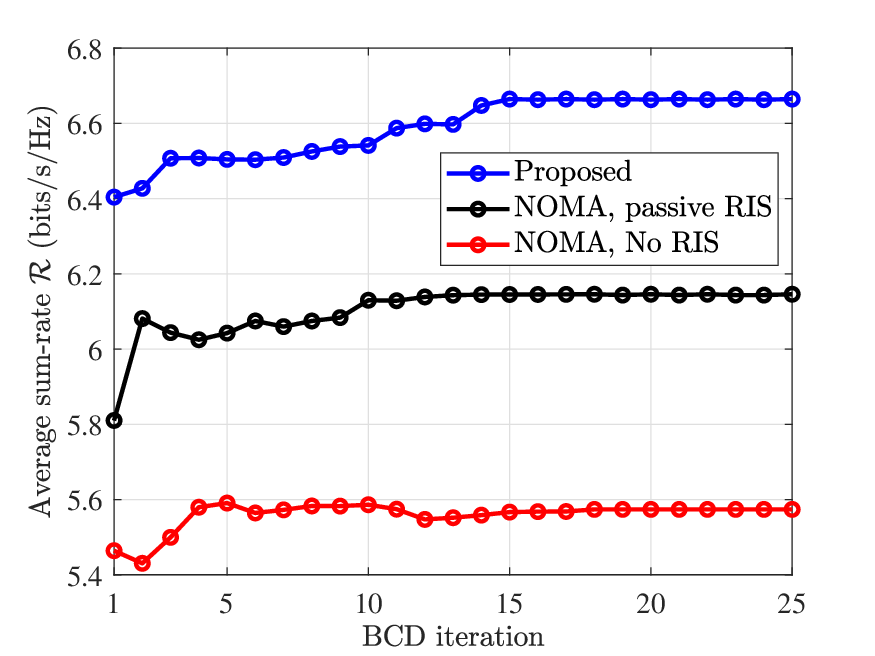}}
\caption{Convergence of proposed \textbf{Algorithm~2}.}
    \label{fig:BCDConv}
\end{subfigure}%
\begin{subfigure}{0.335\textwidth}
\centerline{\includegraphics[scale=0.43]{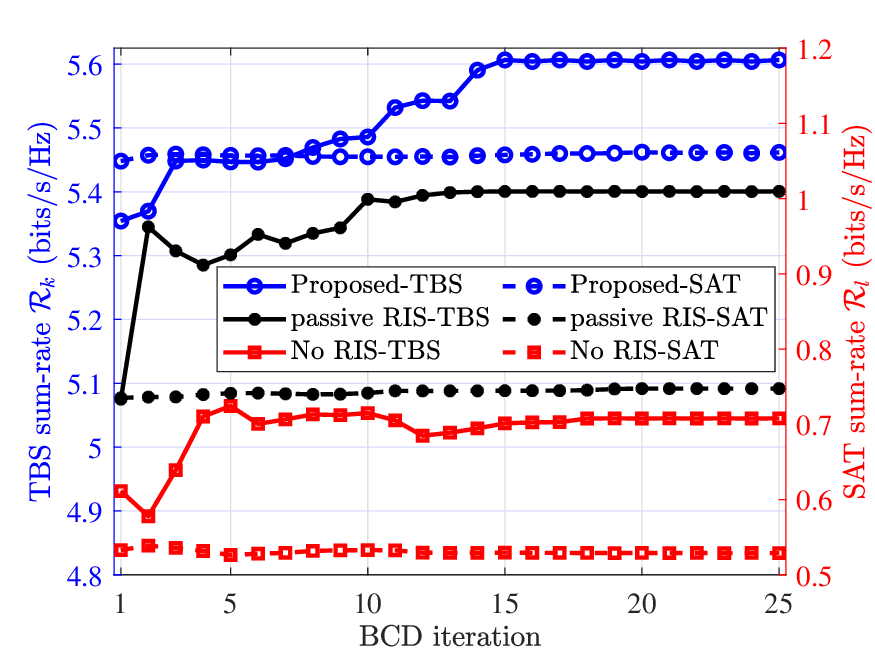}}
\caption{Convergence for $R_k$ and $R_l$.}
    \label{fig:BCDConv_TBS_SAT}
\end{subfigure}%
\begin{subfigure}{0.335\textwidth}
\centerline{\includegraphics[scale=0.43]{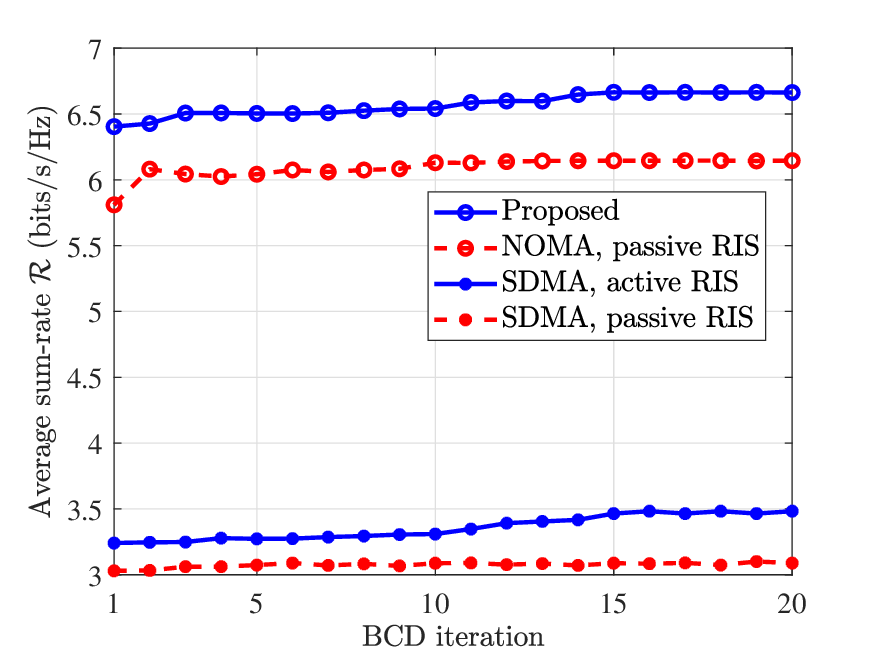}}
\caption{Convergence comparison with SDMA.}
    \label{fig:BCDConv_SDMAComp}
\end{subfigure}
\caption{Convergence of the proposed algorithm, convergence of TBS and SAT user sum-rates, $(R_k)$ \& $(R_l)$, and comparison with the SDMA scheme.}
\label{fig:combined1}
\vspace{-1.25em}
\end{figure*}
%%%%%%%%%%% Figure Section -2 %%%%%%%%
\begin{figure*}[t!]
\centering
\begin{subfigure}{0.335\textwidth}
\centerline{\includegraphics[scale=0.43]{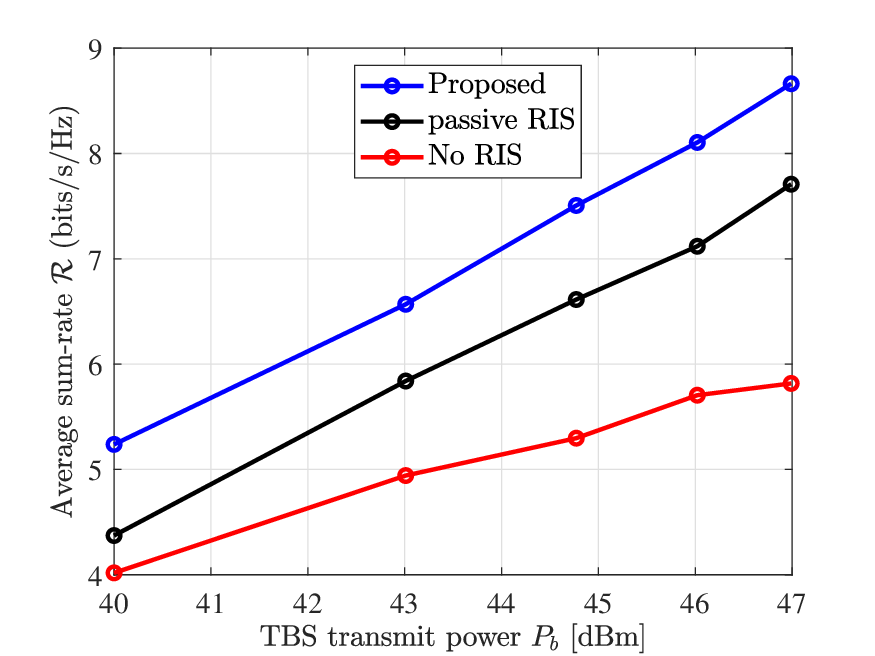}}
\caption{Average sum-rate versus $P_b$ [dBm].}
    \label{fig:Pb}
\end{subfigure}%
\begin{subfigure}{0.335\textwidth} 
\centerline{\includegraphics[scale=0.43] {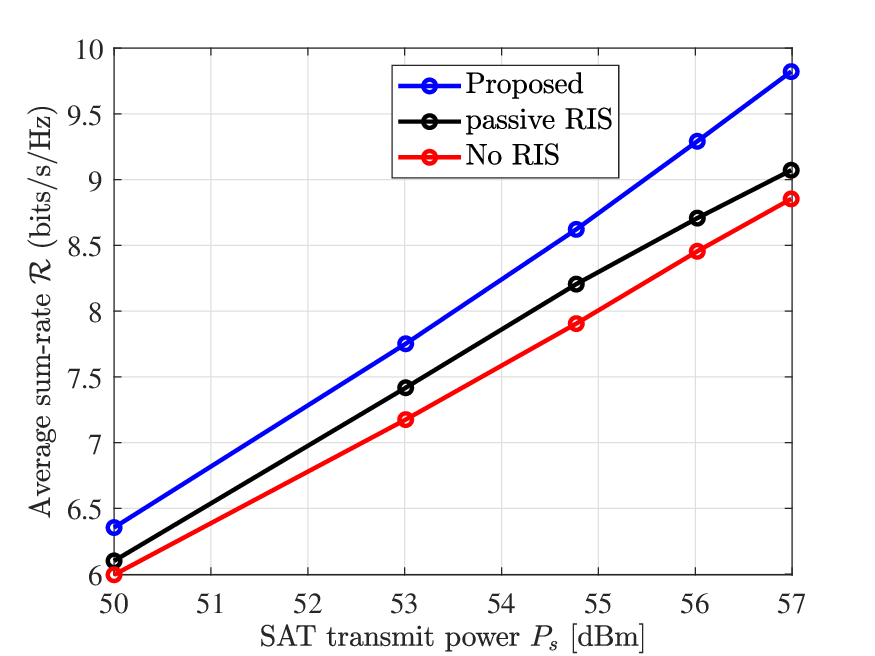}}
\caption{Average sum-rate versus $P_s$ [dBm].}
    \label{fig:Ps}
\end{subfigure}%
\begin{subfigure}{0.335\textwidth}
\centerline{\includegraphics[scale=0.43]{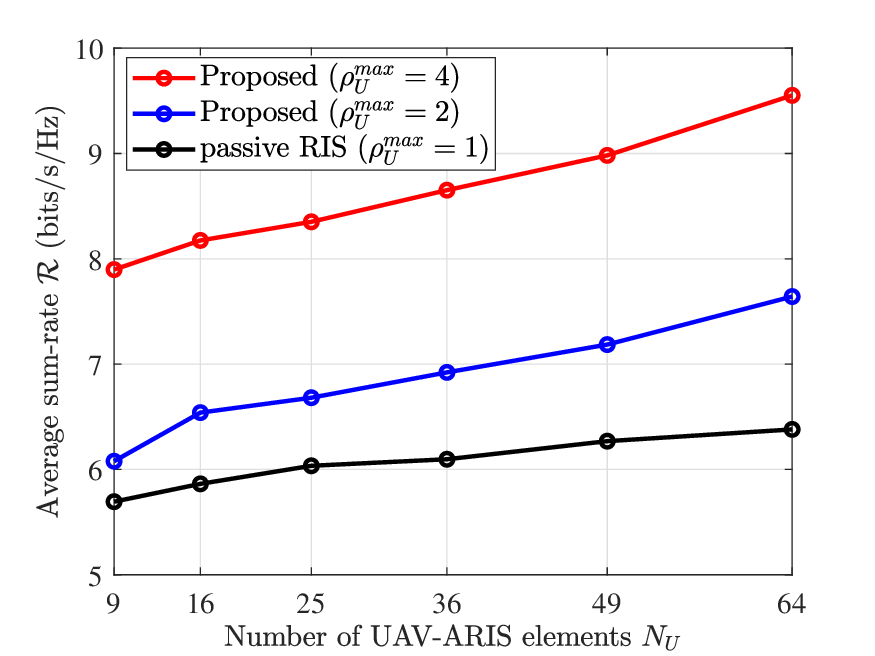}}
\caption{Average sum-rate versus $N_U$.}
    \label{fig:NU}
\end{subfigure}%
\caption{Effect of $P_b$, $P_s$ and $N_U$ on the average sum-rate.}
\label{fig:combined2}
\vspace{-1.25em}
\end{figure*}
\par Fig.~\ref{fig:BCDConv} illustrates the convergence behavior of the proposed BCD-based optimization algorithm in terms of the average sum-rate versus the number of BCD iterations. It is observed that the sum-rate increases monotonically during the initial iterations and converges to a stationary point within approximately $15$ BCD iterations. This monotonic improvement in the sum-rate is guaranteed by the BCD framework, where each subproblem is optimally solved while keeping the remaining variables fixed. This also demonstrates the effectiveness of the BCD algorithm in handling highly coupled optimization variables. In addition, the proposed scheme outperforms all benchmark schemes across the iterations. In particular, it achieves a noticeable performance gain over the \textit{NOMA, passive RIS} scheme due to the additional amplification capability provided by the active RIS. As a result, the proposed scheme achieves an approximately $8.44\%$ higher average sum-rate compared to the passive RIS case. 
\par Fig.~\ref{fig:BCDConv_TBS_SAT} illustrates the convergence behavior of the proposed BCD-based algorithm in terms of average TBS and SAT sum-rates versus the number of BCD iterations. It is observed that the proposed algorithm converges within approximately $15$ BCD iterations for both TBS and SAT sum-rates, indicating the effectiveness of the adopted BCD framework in handling the coupled optimization problem. In addition, the proposed scheme outperforms all benchmark schemes across the iteration range. It is also observed that the TBS sum-rate is higher than that of SAT, mainly due to the more favorable propagation conditions and lower path-loss in terrestrial links compared to satellite channels. In contrast, the \textit{No RIS} scheme has the lowest performance, highlighting the importance of RIS-assisted transmissions in improving spectral efficiency in ITNTN systems.
\par Fig.~\ref{fig:BCDConv_SDMAComp} illustrates the convergence behavior of the proposed BCD-based algorithm in terms of average sum-rate versus the number of BCD iterations for both NOMA and SDMA schemes. The proposed scheme outperforms all considered benchmark schemes. Specifically, the \textit{SDMA, active RIS} scheme achieves about $3.48$ bits/s/Hz, corresponding to a performance degradation of approximately $47.75\%$ compared to the proposed scheme. This shows that NOMA provides better spectral efficiency than SDMA in MU-MISO ITNTN systems. In addition, when comparing active and passive RIS under the same multiple access scheme, active RIS provides approximately $8.44\%$ gain for NOMA and about $12.74\%$ gain for SDMA, highlighting the benefit of signal amplification in improving system performance.
\par Fig.~\ref{fig:Pb} shows the average sum-rate versus the TBS transmit power $P_b$ for the proposed and benchmark schemes. The average sum-rate increases monotonically with $P_b$, as the higher transmit power enhances the received signal strength at users and improves the achievable rate. The proposed scheme outperforms both benchmark schemes across the considered transmit power range. For example, at $P_b=47$ dBm, the proposed scheme achieves a performance gain of  approximately $12.38\%$ over the \textit{passive RIS} benchmark and $48.93\%$ over \textit{No RIS} benchmark. Additionally, when $P_b$ increases from $40$ to $47$ dBm, the proposed scheme achieves an average sum-rate improvement of approximately $65.65\%$, while $P_s$ is fixed at $54.77$ dBm.
\par Fig.~\ref{fig:Ps} shows the average sum-rate versus SAT transmit power $P_s$ for the proposed and benchmark schemes. The average sum-rate increases monotonically with $P_s$, as the higher transmit power enhances the received signal strength at users and improves the achievable rate. The proposed scheme outperforms both benchmark schemes across the considered transmit power range. For example, at $P_s=57$ dBm, the proposed scheme achieves a performance gain of  approximately $8.27\%$ over the \textit{passive RIS} benchmark and $10.94\%$ over \textit{No RIS} benchmark. Additionally, when $P_s$ increases from $50$ to $57$ dBm, the proposed scheme achieves an average sum-rate improvement of approximately $54.63\%$, while $P_b$ is fixed at $43$ dBm.
\par Fig.~\ref{fig:NU} shows the average sum-rate versus the number of UAV-ARIS elements $N_U$ for different maximum UAV-ARIS amplification factors $\rho_{_U}^{max}$. For all schemes, the average sum-rate increases monotonically with $N_U$, as a larger number of reflecting elements enhances channel reconfigurability. In addition, the proposed scheme performs better when the amplification factor is higher. For example, at $N_U = 64$, the proposed scheme with $\rho_{_U}^{max} = 4$ achieves approximately $9.552$ bits/s/Hz, compared to $7.642$ bits/s/Hz for $\rho_{_U}^{max} = 2$ and $6.38$ bits/s/Hz for the passive RIS case ($\rho_{_U}^{max} = 1$). This corresponds to performance gains of about $25\%$ and $49.72\%$, respectively. Moreover, the performance gap between different amplification factors becomes larger as $N_U$ increases. This shows that the benefit of active RIS amplification scales with $N_U$.

%%%%%%%%%%% Figure Section -3 %%%%%%%%
\begin{figure*}[t!]
\centering
\begin{subfigure}{0.335\textwidth} 
\centerline{\includegraphics[scale=0.43] {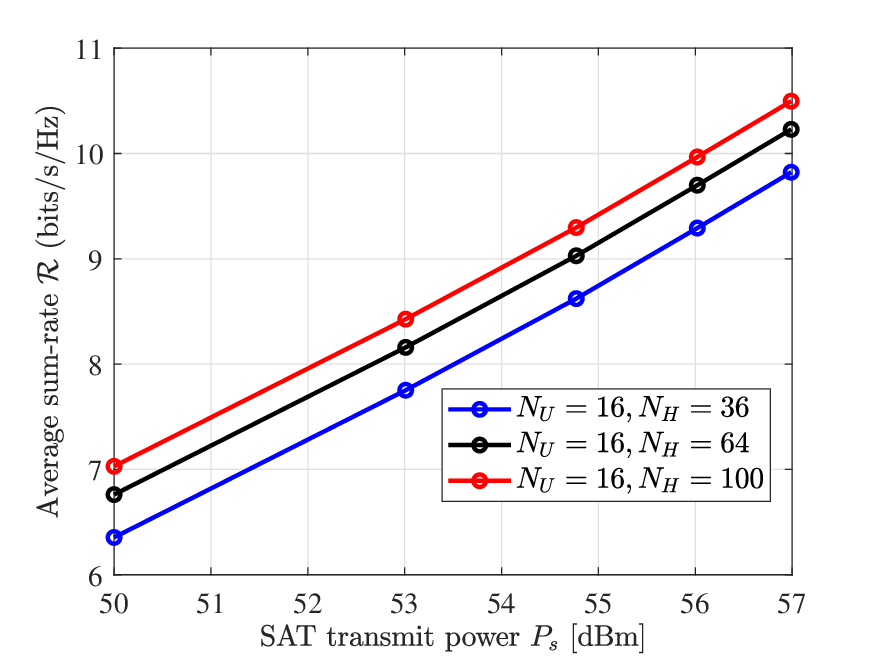}}
\caption{$\mathcal{R}$ versus $P_s$ [dBm] for different $N_H$.}
   \label{fig:SRvsPs_diff_NH}
\end{subfigure}%
\begin{subfigure}{0.335\textwidth}
\centerline{\includegraphics[scale=0.43]{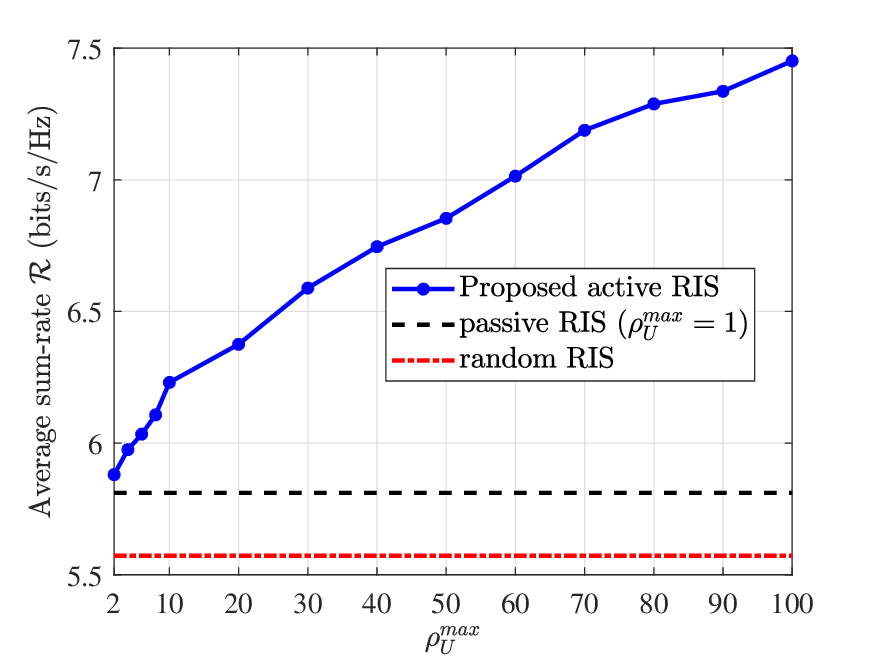}}
\caption{Average sum-rate versus $\rho_{_U}^{max}$.}
    \label{fig:UAV_AMP}
\end{subfigure}%
\begin{subfigure}{0.335\textwidth}
\centerline{\includegraphics[scale=0.43]{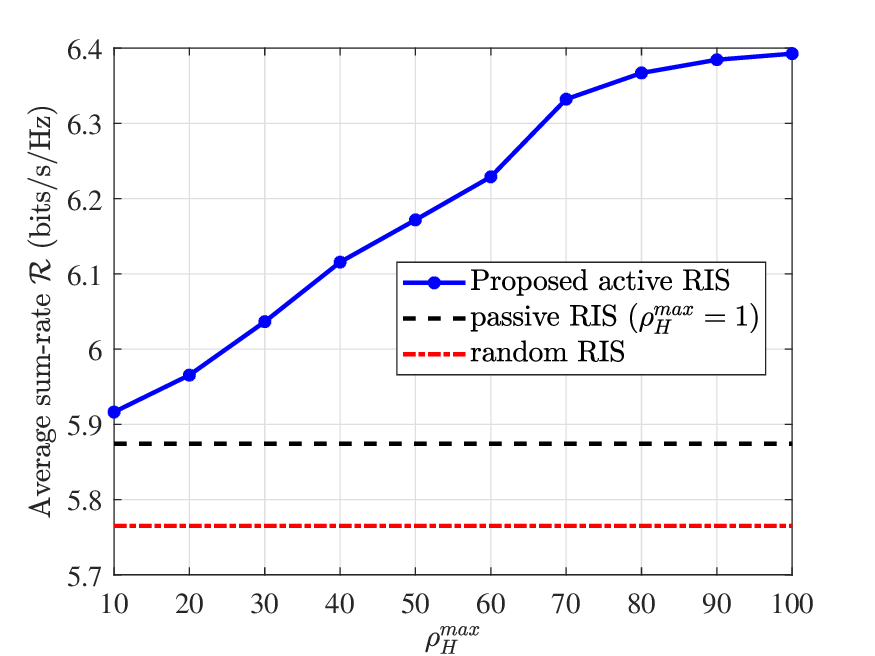}}
\caption{Average sum-rate versus $\rho_{_H}^{max}$.}
    \label{fig:HAP_AMP}
\end{subfigure}
\caption{Effect of $\rho_{_U}^{max}$ and $\rho_{_H}^{max}$ on the average sum-rate.}
\label{fig:combined3}
\vspace{-1.25em}
\end{figure*} 
%%%%%%%%%%% Figure Section -4 %%%%%%%%
\begin{figure*}[t!]
\centering
\begin{subfigure}{0.335\textwidth} 
\centerline{\includegraphics[scale=0.43] {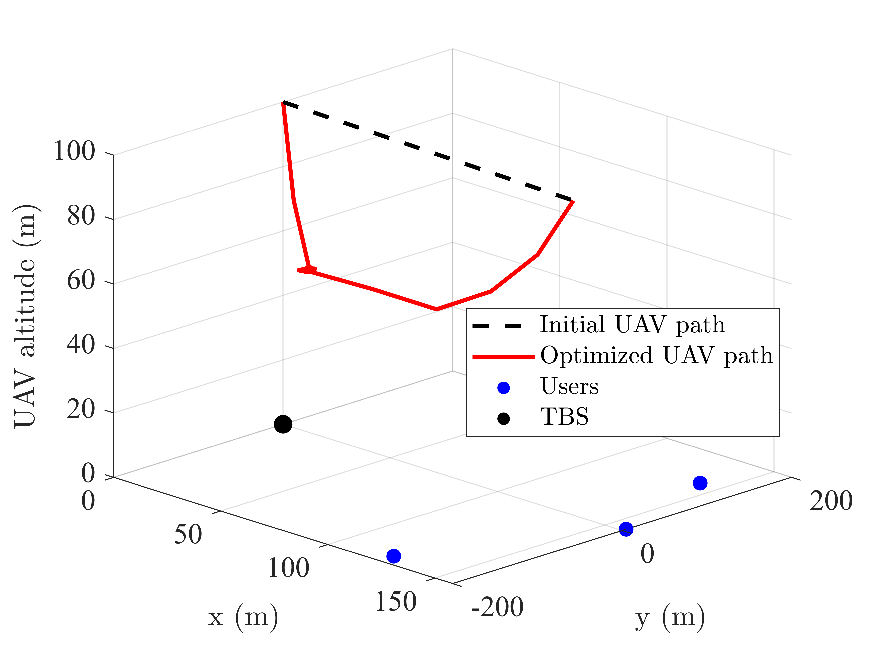}}
\caption{Optimal UAV trajectory.}
    \label{fig:UAV_Traj}
\end{subfigure}%
\begin{subfigure}{0.335\textwidth}
\centerline{\includegraphics[scale=0.43]{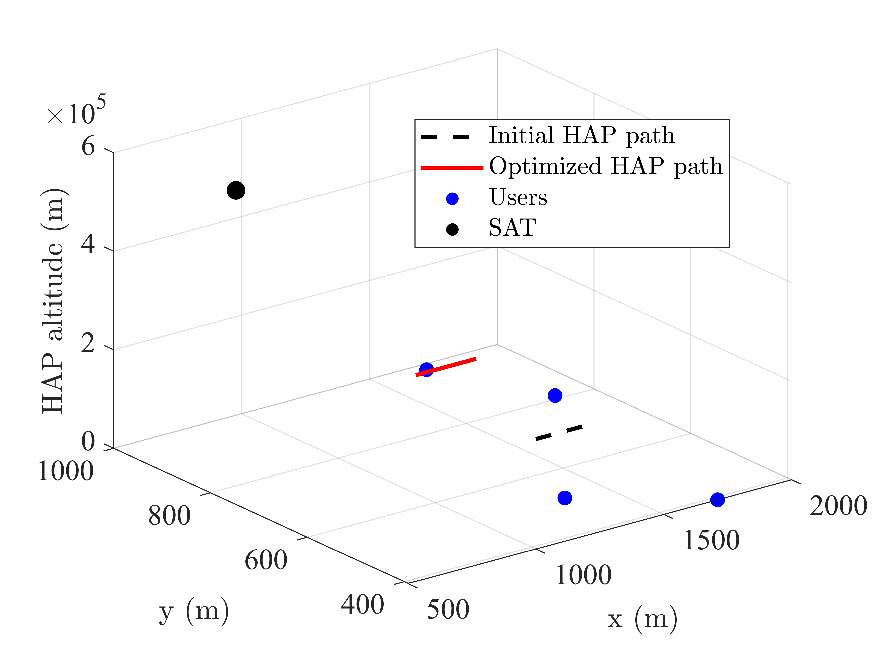}}
\caption{Optimal HAP trajectory.}
    \label{fig:HAP_Traj}
\end{subfigure}%
\begin{subfigure}{0.355\textwidth}
\centerline{\includegraphics[scale=0.43]{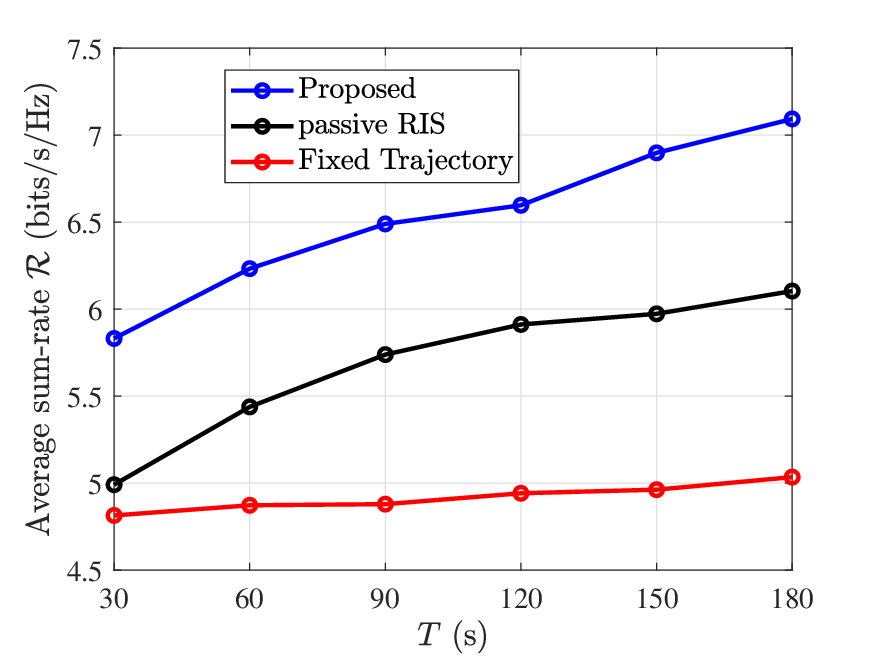}}
\caption{Average sum-rate versus $T$ (s).}
    \label{fig:SRvsT}
\end{subfigure}
\caption{Optimal UAV and HAP trajectories, effect of flight duration $T$ on the average sum-rate}
\label{fig:combined4}
\vspace{-1.25em}
\end{figure*}
\par Fig.~\ref{fig:SRvsPs_diff_NH} illustrates the average sum-rate versus $P_s$ for different number of HAP-ARIS elements $N_H$. It is observed that the average sum-rate increases monotonically with $P_s$ for all cases, due to the higher received signal strength at the user. In addition, increasing $N_H$ leads to consistent performance improvement. Specifically, a larger $N_H$ provides higher passive beamforming gain, which enhances the reflected signal strength and improves the overall channel conditions. As a result, the case with $N_H = 100$ achieves the highest sum-rate followed by $N_H = 64$ and $N_H = 36$. This shows that increasing the RIS size provides noticeable performance gains.
\par Fig.~\ref{fig:UAV_AMP} shows the average sum-rate versus the maximum UAV-ARIS amplification factor $\rho_{_U}^{max}$. It is observed that the average sum-rate of the proposed scheme increases monotonically with $\rho_{_U}^{max}$, as higher amplification enhances the reflected signal strength and improves the effective channel gains. In contrast, the \textit{passive RIS} and \textit{random RIS } benchmark schemes remain unchanged with respect to $\rho_{_U}^{max}$, as they are independent of amplification factor. For example, at $\rho_{_U}^{max}=100$, the proposed scheme achieves a performance gain of  approximately $28.26\%$ over the \textit{passive RIS} benchmark and $33.78\%$ over \textit{random RIS} benchmark. In addition, the average sum-rate shows diminishing returns at larger $\rho_{_U}^{max}$. This is because increasing the amplification factor also amplifies the thermal noise at the RIS, which limits further performance improvement. 
\par Fig.~\ref{fig:HAP_AMP} shows the average sum-rate versus the maximum HAP-ARIS amplification factor $\rho_{_H}^{max}$. It is observed that the average sum-rate of the proposed scheme increases monotonically with $\rho_{_H}^{max}$, as higher amplification enhances the reflected signal strength and improves the effective channel gains. In contrast, the \textit{passive RIS} and \textit{random RIS } benchmark schemes remain unchanged with respect to $\rho_{_H}^{max}$, as they are independent of amplification factor. For example, at $\rho_{_H}^{max}=100$, the proposed scheme achieves a performance gain of  approximately $8.8\%$ over the \textit{passive RIS} benchmark and $10.89\%$ over \textit{random RIS} benchmark. In addition, the average sum-rate shows diminishing returns at larger $\rho_{_H}^{max}$. This is because increasing the amplification factor also amplifies the thermal noise at the RIS. 
\par Fig.~\ref{fig:UAV_Traj} shows the optimized 3D UAV trajectory compared with the initial UAV path, along with the positions of the TBS and terrestrial users. It is observed that the optimized trajectory deviates from the straight-line initialization, indicating that the trajectory optimization adapts to the spatial distribution of users. Specifically, the UAV dynamically adjusts both its horizontal position and altitude to better serve the user while maintaining connectivity with the TBS. The UAV tends to move closer to the users to enhance channel conditions and improve received signal strength. In addition, the altitude variation along the trajectory reflects a trade-off between path loss reduction and maintaining favorable propagation conditions. As a result, this adaptive 3D trajectory design further improves the average sum-rate.
\par Fig.~\ref{fig:HAP_Traj} depicts the optimized 3D HAP trajectory compared with the initial path, along with the positions of the SAT and satellite users. The optimized trajectory deviates from the initial path, indicating that the trajectory design adapts to the spatial distribution of the users and the satellite link. Specifically, the HAP adjusts its horizontal position to better align with the clustered users, improving the channel conditions. Compared to the UAV case, the HAP operates at a higher altitude, resulting in a smoother trajectory with less variation. This is because the HAP primarily acts as a high-altitude relay between the satellite and users, where large-scale path loss is dominated and abrupt movements provide limited additional gain. In addition, the optimized trajectory balances the trade-off between maintaining a strong satellite-HAP link and enhancing HAP-user channels. As a result, the HAP stays within a favorable region rather than aggressively moving towards the users.
\par Fig.~\ref{fig:SRvsT} illustrates the average sum-rate versus flight duration $T$. It is observed that the average sum-rate increases with $T$ for all schemes. This is because a longer flight duration provides greater flexibility for UAV/HAP trajectory optimization, allowing aerial platforms to better adjust their positions to improve channel conditions. The proposed scheme outperforms the benchmark schemes across all values of $T$. For example, at $T = 180$ s, the proposed scheme achieves approximately $7.0932$ bits/s/Hz, compared to about $6.1038$ bits/s/Hz for passive RIS and $5.0343$ bits/s/Hz for the fixed trajectory scheme, corresponding to performance gains of approximately $16.2\%$ and $40.89\%$, respectively. In addition, the performance improvement shows diminishing returns as $T$ increases. This behavior is due to the fact that, under practical mobility constraints, the UAV/HAP can already reach near-optimal positions, and additional flight time provides only limited further improvement.
\section{Conclusion}\label{ref:Conclusion}
This paper investigated a dual-aerial ARIS-assisted NOMA-based ITNTN, where a UAV-ARIS and a HAP-ARIS simultaneously assist communications between terrestrial and satellite networks serving terrestrial and satellite users. We formulated an average sum-rate maximization problem by jointly optimizing transmit beamforming, ARIS coefficients, and 3D trajectories of the UAV and HAP under practical constraints. To address the resulting highly non-convex problem, a BCD-based framework was developed, integrating WMMSE for beamforming, manifold-based RCG for RIS phase-shifts, SCA for ARIS amplification factors, and first-order Taylor approximations for trajectory design. The simulation results demonstrated that the proposed joint design achieved significant performance gains over benchmark schemes, highlighting the effectiveness of dual-aerial ARIS deployment and joint communication–mobility optimization in improving spectral efficiency. Future work will consider more practical scenarios, including imperfect channel state information, energy efficiency optimization, and multi-agent learning-based real-time adaptive resource allocation.

\bibliographystyle{IEEEtran}   
\bibliography{ref} 
\end{document}